\documentclass[letterpaper,twocolumn,10pt]{article}
\usepackage{usenix}

\usepackage{cite}
\usepackage{amsmath,amssymb,amsfonts, amsthm}
\usepackage{framed}

\usepackage{graphicx}
\usepackage{textcomp}
\usepackage{xcolor}
\usepackage{booktabs}
\usepackage{xspace} %
\usepackage{array} %
\usepackage{paralist} %
\usepackage{dirtytalk} %
\usepackage{wasysym} %
\usepackage{enumitem}

\theoremstyle{definition}

\newtheoremstyle{italicbody}
  {}                %
  {}                %
  {\itshape}        %
  { }                %
  { \bfseries}       %
  {:}               %
  { }               %
  {}                %

\theoremstyle{italicbody}

\newtheorem{definition}{Definition}[section]
\newtheorem{theorem}{Theorem}[section]

\usepackage{algorithm}
\usepackage[noend]{algpseudocode}

\usepackage[acronym]{glossaries}

\usepackage{makecell}
\setcellgapes{0pt}

\usepackage{cryptocode}

\usepackage{caption}
\usepackage{subcaption}

\usepackage{authblk} %
 
\usepackage{soul} %
\usepackage{xcolor}
\definecolor{lightblue}{HTML}{50C878}
\sethlcolor{lightblue}

\newcommand{\covered}{\checkmark \xspace}
\newcommand{\notCovered}{\texttimes \xspace}

\newcommand{\epoch}{epoch\xspace}

\newcommand{\epochs}{epochs\xspace}

\newcommand{\ecounter}{$\mathsf{e}$\xspace}
\newcommand{\ecounteri}[1]{$\mathsf{e}_{#1}$\xspace}

\newcommand{\ecounters}{$\mathsf{epochs}$\xspace}

\newcommand{\sk}{$\mathsf{sk}$\xspace}
\newcommand{\pk}{$\mathsf{pk}$\xspace}
\newcommand{\crs}{$\mathsf{crs}$\xspace}
\newcommand{\token}{$\mathsf{token}$\xspace}

\newcommand{\tokens}{tokens\xspace}

\newcommand{\tokenvar}{$\mathsf{token}$\xspace}
\newcommand{\tokenvari}[1]{$\mathsf{token}_{#1}$\xspace}

\newcommand{\tokenvarm}{$\mathsf{token}_m$\xspace}
\newcommand{\tokenvars}{$\mathsf{tokens}$\xspace}

\newcommand{\seed}{$\mathsf{seed}$\xspace}

\newcommand{\duration}{duration\xspace}
\newcommand{\durationvar}{$\mathsf{dur}$\xspace}

\newcommand{\tsInit}{$\mathsf{ts}_0$\xspace}

\newcommand{\challenge}{$\mathsf{challenge}$\xspace}

\newcommand{\nonce}{$\mathsf{nonce}$\xspace}

\newcommand{\GS}{$\mathsf{blacklist_{iss}}$\xspace}
\newcommand{\GSNew}{$\mathsf{blacklist'_{iss}}$\xspace}
\newcommand{\RevList}{$\mathsf{RevList_{iss}}$\xspace}
\newcommand{\RevListNew}{$\mathsf{RevList_{iss}}'$\xspace}

\newcommand{\VC}{VC\xspace}
\newcommand{\VCs}{VCs\xspace}
\newcommand{\vc}{vc\xspace}

\newcommand{\VP}{VP\xspace}
\newcommand{\VPs}{VPs\xspace}
\newcommand{\vp}{vp\xspace}

\newcommand{\setup}{setup\xspace}
\newcommand{\Setup}{Setup\xspace}
\newcommand{\issuance}{issuance\xspace}
\newcommand{\Issuance}{Issuance\xspace}
\newcommand{\refresh}{refresh\xspace}
\newcommand{\Refresh}{Refresh\xspace}
\newcommand{\revocation}{revocation\xspace}
\newcommand{\Revocation}{Revocation\xspace}
\newcommand{\presentation}{presentation\xspace}

\newcommand{\Presentation}{Presentation\xspace}
\newcommand{\verification}{verification\xspace}
\newcommand{\Verification}{Verification\xspace}

\newcommand{\name}{\emph{zkToken}\xspace}

\newcommand{\oracle}[1]{$\mathcal{O}^{\mathsf{#1}}$\xspace}

\newcommand{\expirationperiod}{expiration period\xspace}

\newcommand{\verificationperiod}{verification period\xspace}

\newcommand{\vcspec}{W3C Verifiable Credentials Specification\xspace}

\newcommand{\iotaid}{IOTA Identity\xspace}
\newcommand{\irma}{IRMA\xspace}

\newcommand{\validuntil}{$\mathsf{exp}$\xspace}

\newcommand{\claims}{$\mathsf{claims}$\xspace}
\newcommand{\m}{$\mathsf{m}$\xspace}

\newcommand{\h}{$\mathsf{h}$\xspace}
\newcommand{\x}{$x$\xspace}
\newcommand{\w}{$\mathsf{w}$\xspace}
\newcommand{\sig}{$\mathsf{sig}$\xspace}

\usepackage{graphicx} %
\title{\Large \bf zkToken: Empowering Holders to Limit Revocation Checks for Verifiable Credentials}%

\author[1]{Praveensankar Manimaran}
\author[1]{Mayank Raikwar}
\author[1]{Thiago Garrett}
\author[2]{Arlindo F.\ da Conceição}
\author[3]{Leander Jehl}
\author[1]{Roman Vitenberg}
\affil[1]{University of Oslo, Norway}
\affil[2]{Federal University of São Paulo, Brazil}
\affil[3]{University of Stavanger, Norway}

\microtypecontext{spacing=nonfrench}
\begin{document}

\maketitle

\begin{abstract}
	Systems managing Verifiable Credentials are becoming increasingly popular. Unfortunately, their support for revoking previously issued credentials allows verifiers to effectively monitor the validity of the credentials, which is sensitive information. While the issue started to gain recognition, no adequate solution has been proposed so far.
	
	In this work, we propose a novel framework for time-limited continuous verification. The holder is able to individually configure the verification period when sharing information with the verifier, and the system guarantees proven untraceability of the revocation status after the verification period expires. Differently from existing systems, the implementation adopts a more scalable blacklist approach where tokens corresponding to revoked credentials are stored in the registry. The approach employs ZK proofs that allow holders to prove non-membership in the blacklist. In addition to theoretically proving security, we evaluate the approach analytically and experimentally and show that it significantly improves bandwidth consumption on the holder while being on par with state-of-the-art solutions with respect to the other performance metrics.

\end{abstract}

\section{Introduction}

Verifiable Credentials (\VCs)~\cite{sporny2021w3cvc, openidvc, mazzocca2025survey, tan2023survey, hesse2023credentials, blazy2023credentials} represent conventional credentials, such as passports and driving licenses, in a digital format. 
Similar to how individuals store and share conventional credentials, digital identity holders store \VCs and share them with verifiers. 
\VC management systems~\cite{irma2024revocation, anoncreds2024} 
are increasingly adopted in the industry, including both public~\cite{ebsivc, orgbook} and private~\cite{dock, entra, indicio, digitalbazaar} sector organizations. This adoption is further fuelled by commonly accepted standards such as the  \vcspec~\cite{sporny2021w3cvc}.

A \VC can be revoked for reasons such as misuse or when the corresponding holder loses access to the \VC. 
For example, an employee ID issued in the form of a VC will be revoked when the corresponding employee leaves the organization. 
Therefore, it is critical for verifiers to ascertain that the \VC has not been revoked prior to granting services. 
Support for revocation is provided by many systems, including IRMA~\cite{irma2024revocation}, Anoncreds~\cite{anoncreds2024}, and  \iotaid~\cite{revocationbitmap2022, revocationtimeframe2024}.

Unfortunately, revocation introduces a new issue: the verifier may now be able to repeatedly probe the revocation status and thus, effectively monitor the validity of the \VC. 
A \VC may pertain to sensitive information, such as employee's position with the current employer, credit score, or healthcare-related situation. 
A holder may want the verifier to perform the verification within a limited time after sharing the \VC rather than indefinitely. 
This issue is considered one of the key privacy challenges by EBSI~\cite{ebsivc,ebsi2023revocationwhitepaper}, an initiative of the European Commission and the European Blockchain Partnership. 
The issue is also recognized by~\cite{revocationtimeframe2024} and~\cite{Sitouah2024untrace}. IRMA, Anoncreds, and the academic works of \cite{Sitouah2024untrace} and \cite{rosenberg2023zkcreds} allow the verification period to be shorter than the \VC expiration period, see Table~\ref{tab:related_work}. Unfortunately, they do not let the holder configure the verification period, thereby lending themselves only to specific use cases. Adding this configurability to existing approaches is challenging for two reasons: (a) the main responsibility of cryptographic computations in existing approaches falls on the issuer creating the \VC rather than on the holder sharing it, and (b) existing approaches rely on state compaction techniques such as accumulators~\cite{loporchio2023survey} to achieve good performance. Alas, these techniques prevent implementation of configurable verification period.

The main contribution of our work is that we propose a novel framework for time-limited continuous verification. 
The holder is able to individually configure the verification period when sharing information with the verifier, and the system guarantees proven \textbf{untraceability} of the revocation status after the verification period expires. 
The framework includes two additional requirements: (a) one-time sharing for a continuous verification: the verifier does not need to contact the holder after the initial sharing is completed, and (b) unobservable verification: the issuer should not be able to learn anything about the verification. 
We provide a scalable solution for this framework, and we demonstrate that all cryptographic operations take practically acceptable times.

Our solution is a radical departure from existing schemes: all previously known solutions for \VC revocation are based on a whitelist of tokens corresponding to issued unrevoked \VCs. In contrast, we employ a blacklist of previously revoked \VCs combined with ZK proofs that allow holders to prove non-membership in the blacklist.
Since in practice, the number of revoked \VCs during a period of time is several times smaller than the number of issued \VCs during the same period, the blacklist approach is inherently more efficient in terms of the bandwidth consumption and storage. For example, if we consider the specific domain of Web’s PKI, only $1$\% certificates are revoked~\cite{schanck2025clubcards, liu2015revocation}.  To compensate, many systems based on the whitelist approach employ state compaction techniques, such as accumulators. Accumulators are even better than blacklist in terms of space complexity: they achieve constant complexity compared to the linear complexity of blacklist or whitelist without accumulators. However, state compaction techniques do not mitigate bandwidth consumption and besides, they prevent implementation of configurable verification period.

In summary, our contributions are as follows:
\begin{asparaitem}
	\item We introduce a novel framework for time-limited continuous verification, with formally defined requirements.
	\item We propose a blacklist-based solution that satisfies the requirements.
	\item We prove the correctness of our protocol under standard cryptographic assumptions.
	\item We evaluate the proposed solution both analytically and experimentally. We conclude that it significantly improves bandwidth consumption on the holder while providing a nuanced comparison with state-of-the-art solutions with respect to the other performance metrics. %
\end{asparaitem}

\begin{table*}[ht!]
\caption{Related Work (\covered - covered; \notCovered - not covered;)}
\label{tab:related_work}
\begin{center}
\newcolumntype{x}[1]{>{\centering\arraybackslash}m{#1}}
\begin{tabular}{x{5cm}x{4cm}x{2cm}x{2cm}x{2cm}}
\toprule

 & \multicolumn{4}{c}{\textbf{Properties}} \\
\textbf{Related Work} & Separation between expiration and verification  & Configurable Period & One-time VP sharing   & Unobservable verification \\
\midrule

IRMA~\cite{irma2024revocation}, Anoncreds~\cite{anoncreds2024}, Sitouah~\textit{et al.}~\cite{Sitouah2024untrace}, zk-creds~\cite{rosenberg2023zkcreds} & \covered & \notCovered & \notCovered  &  \covered  \\
\midrule

EVOKE~\cite{carlo2024evoke} & \notCovered & \notCovered  & \notCovered & \covered   \\
\midrule

Prevoke~\cite{praveen2024prevoke} & \notCovered & \notCovered  & \notCovered & \covered   \\
\midrule

Schumm~\textit{et al.}~\cite{schumm2023revocation}, \iotaid: Revocation Bitmap 2022~\cite{revocationbitmap2022} & \notCovered & \notCovered  & \covered & \covered    \\
\midrule

 \iotaid: Revocation Timeframe 2024~\cite{revocationtimeframe2024} & \notCovered & \notCovered & \notCovered  & \covered     \\
\midrule

Papathanasiou~\textit{et al.}~\cite{Papathanasiou2024revocation} & \notCovered &  \notCovered & \covered  &   \notCovered \\
\midrule
This work & \covered & \covered  & \covered   & \covered  \\
\bottomrule
\end{tabular}

\end{center}
\end{table*}

\section{Related Work}

\label{sec:related_work}

In this section, we show how \name differs from other systems that support revocation of verifiable credentials. 
Our description focuses on the coverage of requirements as summarized in Table~\ref{tab:related_work} and the reasons explaining why it is not trivial for the systems to satisfy the missing requirements. 
First, we present the systems that separate \expirationperiod of \VCs and \verificationperiod of VPs.  
Subsequently, we discuss other implementations supporting \VC revocation.
We do not cover protocols designed specifically for X.509 certificates such as Certificate Revocation List (CRL)~\cite{CRL_rfc} and Online Certificate Status Protocol (OCSP)~\cite{OCSP_rfc}. 
These protocols are not suitable for systems based on verifiable credentials due to differences in design goals and functionalities.

All works except Papathanasiou~\textit{et al.}~\cite{Papathanasiou2024revocation} satisfy unobservable verification
since they implement registry using DLTs with public read-access. 
Essentially, proofs needed for verification are stored in DLTs and accessed by verifiers. 
Therefore, issuers do not observe anything about the verification process.

The following systems in the related work have separate \verificationperiod for VPs: (a) IRMA~\cite{irma2024revocation, baldimtsi2017revocation}, (b) Anoncreds~\cite{anoncreds2024,camenisch2009revocation}, (c) Sitouah~\textit{et al.}~\cite{Sitouah2024untrace}, and (d) zk-creds~\cite{rosenberg2023zkcreds}.
These systems utilize accumulators~\cite{loporchio2023survey} and Zero-Knowledge Proofs~\cite{de1992zero, groth2016zkp, chiesa2020marlin} for the purpose of revocation check. 
At a high level, all these systems follow a whitelist approach, where identifiers corresponding to valid VCs are inserted in the accumulator during \issuance and removed during \revocation. 
To prove non-revocation, a holder generates a witness and computes a ZK proof. 
The ZK proof proves the inclusion of identifiers in the accumulator. 
In these systems, the \verificationperiod for VPs corresponds to the duration between accumulator updates.
Therefore, the \verificationperiod is derived rather than controlled. 
Due to the witness update problem~\cite{camenisch2009revocation} in accumulators, a holder needs to update the witness to account for the removed identifiers during \presentation.
Therefore, it is not possible for the holder (a) to configure the \verificationperiod since a holder needs to update the witness every time the accumulator changes, and (b) to share witnesses that are valid in the future since new witnesses need to be recomputed every time the accumulator changes.
Therefore, these systems do not satisfy configurable period or one-time VP sharing.

Dynamic Status List~\cite{ebsi2024bloomfilter} is a protocol by EBSI~\cite{ebsivc} designed to perform untraceable revocation status check. 
This protocol offers separation between \expirationperiod and \verificationperiod. 
However, it has a security issue since holders can present fake \tokens to verifiers. The issue has been acknowledged by EBSI; it is nontrivial to fix without significant changes to the protocol~\cite{ebsi-proposal}.
Due to this, we do not include the protocol in Table~\ref{tab:related_work}.

Next, we discuss the works that do not offer separation between \expirationperiod and \verificationperiod.
EVOKE~\cite{carlo2024evoke} proposes a revocation protocol that is based on accumulators. 
EVOKE does not use ZKPs unlike the other-based accumulator-based works.
Holders directly shared accumulator witnesses in plain-text. 
Once a verifier gets access to a VP consisting of a witness, the verifier itself can update the witness and continously monitor the revocation status.
Due to the witness update issue~\cite{camenisch2009revocation} in accumulators, this protocol does not satisfy the one-time VP sharing requirement.

Prevoke~\cite{praveen2024prevoke} proposes a  protocol that combines a bloom filter and a merkle tree accumulator~\cite{loporchio2023survey} to store and verify the revocation status of \VCs. 
A holder shares the bloom filter index and merkle witness with the verifier. 
This protocol does not address one-time VP sharing since holders need to present an updated Merkle witnesses.

Schumm \textit{et al.}~\cite{schumm2023revocation} and Revocation Bitmap 2022~\cite{revocationbitmap2022} uses bloom filters to store revoked \VCs. 
A holder shares bloom filter index encoded in its \VC with a verifier.
The verifier checks index inclusion in the bloom filter to verify non-revocation.
The protocol addresess one-time VP sharing since holders are not required to provide any proofs after bloom filter indexes have been shared.

Revocation Timeframe 2024~\cite{revocationtimeframe2024} extends Revocation Bitmap 2022~\cite{revocationbitmap2022} to address the problem of untraceability. 
The extension consists of an issuer periodically re-issuing a \VC.
The issuer is responsible for setting the expiry period, which impacts how long a verifier can track the revocation and how often the issuer needs to re-issue VCs. 
This technique neither separates the \expirationperiod and \verificationperiod nor addresses the configurable period and one-time VP sharing requirements.

Papathanasiou~\textit{et al.}~\cite{Papathanasiou2024revocation} uses a hashtable to store revoked VCs. 
For each VC, a unique identifier based on VRF proofs are inserted in the hashtable.
Holders share the identifiers with verifiers. 
The protocol addresses one-time VP sharing since no other proofs are needed once a VP is shared.
It does not address unobservable verification since the hashtables are managed by issuers rather than DLTs.

\section{Cryptographic Preliminaries}
\label{sec:background_crypto}

\subsection{Zero-Knowlege Proofs}

Zero-Knowledge Proof (ZKP) schemes~\cite{de1992zero, groth2016zkp, chiesa2020marlin} allow a prover to convince a verifier of a statement’s validity without revealing any additional information.  

Formally, a ZKP system consists of a prover $P$ and a verifier $V$. Given an efficiently computable binary relation $R(x, w)$, where $x$ is a statement and $w$ is a witness, the corresponding language is $\mathcal{L} = \{ x \mid \exists w \text{ s.t. } R(x, w) = 1 \}$. A ZKP scheme ensures that $P$ can prove $x \in \mathcal{L}$ without disclosing $w$.

\noindent \textbf{Properties.}
A Zero-Knowledge Proof system satisfies the following three properties:
\begin{itemize}[noitemsep, topsep=0pt]
\item \textit{Completeness:}  If $x\in \mathcal{L}$, Prover $P$ can convince the verifier $V$ that $P$ has the correct input.
\item \textit{Soundness:} If $x \notin \mathcal{L}$, the statement $x$ cannot be proven.
\item \textit{Zero-Knowledge:} Verifier $V$ learns nothing beyond the truth of statement $x$.
\end{itemize}

We use Groth16~\cite{groth2016zkp} ZKP scheme in our solution.   
      Groth16 ZKP scheme consists of the following algorithms:
    \begin{itemize}[noitemsep, ,topsep=0pt]
        \item $\mathsf{G16.Setup}\,(1^{\lambda},\, C) \rightarrow crs$: Given a security parameter $\lambda$ and a circuit $C$,  output a common reference string (CRS) $crs$.
        \item $\mathsf{G16.Prove}\,(x,crs,w) \rightarrow \pi $: Given a statement $x$, $crs$, and witness $w$, output a proof $\pi_\mathsf{zkp}$ .
        \item $\mathsf{G16.Verify}\,(x,crs,\pi) \rightarrow \{0, 1\}$: Given a statement $x$, $crs$, and proof $\pi$, output accept or reject.
    \end{itemize}

       The notations used in the algorithms are described in zk-creds~\cite{rosenberg2023zkcreds}.

\paragraph*{Zero-Knowledge Circuit.} A ZK circuit is a mathematical abstraction that encodes a computer program as a set of constraints suitable for Zero-Knowledge Proofs (ZKPs). The execution of the program is represented as a constraint system comprising two types of constraints:

\begin{enumerate}[noitemsep, topsep=0pt]
    \item \textit{Private constraints}, which capture relationships involving private inputs (known only to the prover) and internal variables of the circuit.
    \item \textit{Public constraints}, which involve public inputs (known to both the prover and the verifier) and internal variables.
\end{enumerate}

A valid witness set of input values satisfying all constraints serves as evidence of correct program execution. The prover uses this witness to construct a proof asserting the correctness of the computation without revealing any private input.

In the Groth16 ZKP scheme, once the ZK circuit is defined, a common reference string, crs, is generated based on a given security parameter. This crs is subsequently used in both the proof generation and verification phases, ensuring soundness and succinctness of the proof system.

\section{Preliminaries on Verifiable Credentials}
\label{sec:background_vc}

\begin{figure}
    \centering
    \includegraphics[scale=0.4]{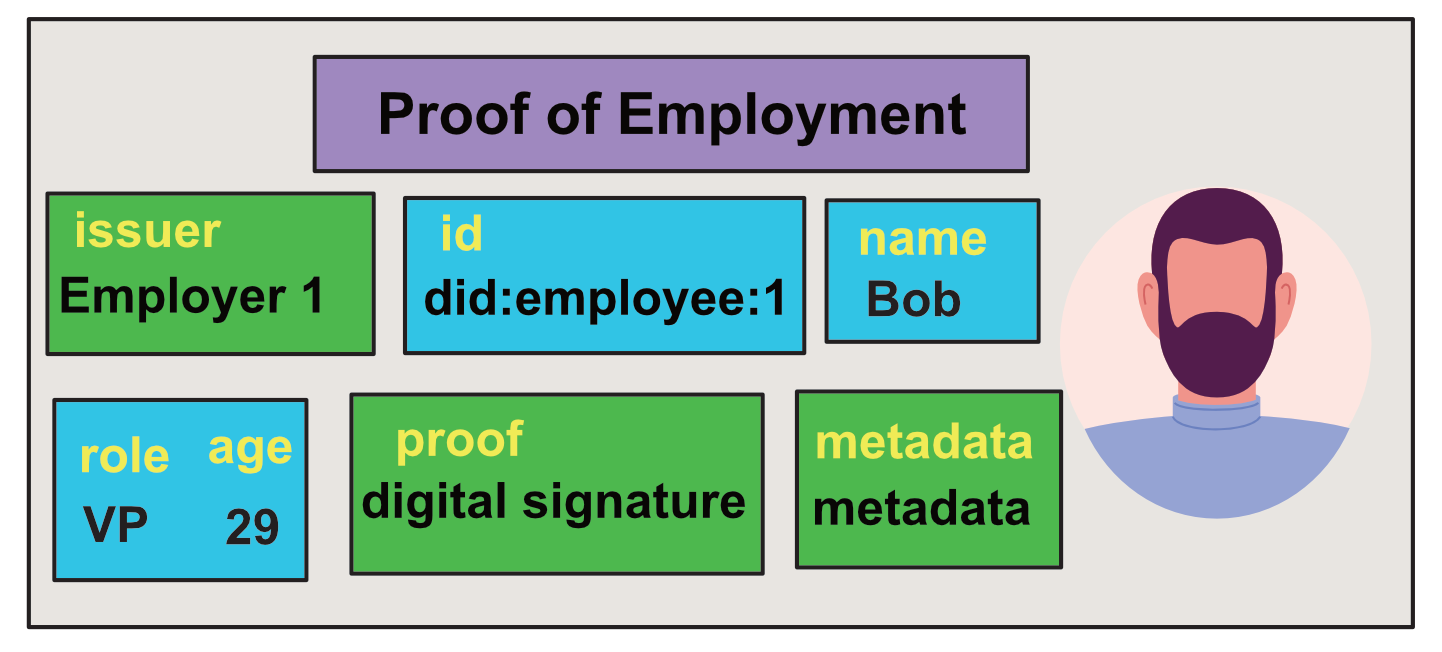}
    \caption{Example of VC: an employee identification.}
    \label{fig:vc_proof_of_employment}
   
\end{figure}

Verifiable Credentials (\VCs)~\cite{sporny2021w3cvc, schardong2022self} are digital representations of conventional credentials, including passports, diplomas, and driving licenses. 
They are designed to be secure, privacy-preserving, and machine-verifiable.
Similar to conventional credentials, a typical \VC consists of information about the issuer (e.g issuer's name) and claims such as name, and age.
An example of a \VC is depicted in Fig.~\ref{fig:vc_proof_of_employment}, showcasing details about an employee covering the name of an employer who issued the \VC and claims about the employee, covering the employee's identifier, name, age, role, and picture.
In addition, \VCs include digital signatures of issuers to verify authenticity and integrity. 
Authenticity stipulates that a \VC is provisioned by a specific issuer. Integrity ensures that the contents of a \VC are not tampered with.

Verifiable Presentations (\VPs) are constructs atop \VCs that provide privacy advantages such as selective disclosure of a subset of claims from a \VC, as well as aggregation of data from multiple \VCs. The \vcspec~\cite{sporny2021w3cvc} introduced \VCs and is currently in the process of standardizing data models for \VCs and \VPs.

\paragraph*{System Model.}

We present the system model introduced in the W3C Verifiable Credentials Specification~\cite{sporny2021w3cvc}. 
The system model is consistent across other state-of-the-art systems~\cite{revocationtimeframe2024, revocationbitmap2022, ebsi2024bloomfilter, anoncreds2024, irma2024revocation} and literature~\cite{carlo2024evoke, rosenberg2023zkcreds, praveen2024prevoke} based on VCs. 
The system model consists of the following entities: a set of issuers, a set of holders, a set of verifiers, and a registry.

\begin{asparadesc}
    \item[Issuer] - An issuer issues VCs to holders.  It also revokes previously issued VCs.  
    
    \item[Holder] - A holder receives a \VC  from an issuer.  Once the VC is received, the holder would store it. The holder also constructs one or more \VPs from the \VC. 
    \item[Verifier] - A verifier receives \VPs from holders. 
    
    The verifier verifies authenticity, integrity, and revocation status of the \VPs.
    \item[Registry] - The registry stores metadata such as issuers' public keys and proofs to assist the verification of VPs. Some of the possible implementations for a registry are DLTs, centralized servers, etc. 
\end{asparadesc}

\paragraph*{Use Case: Workforce Identity.}

\begin{figure*}
    \centering
    \includegraphics[scale=0.6]{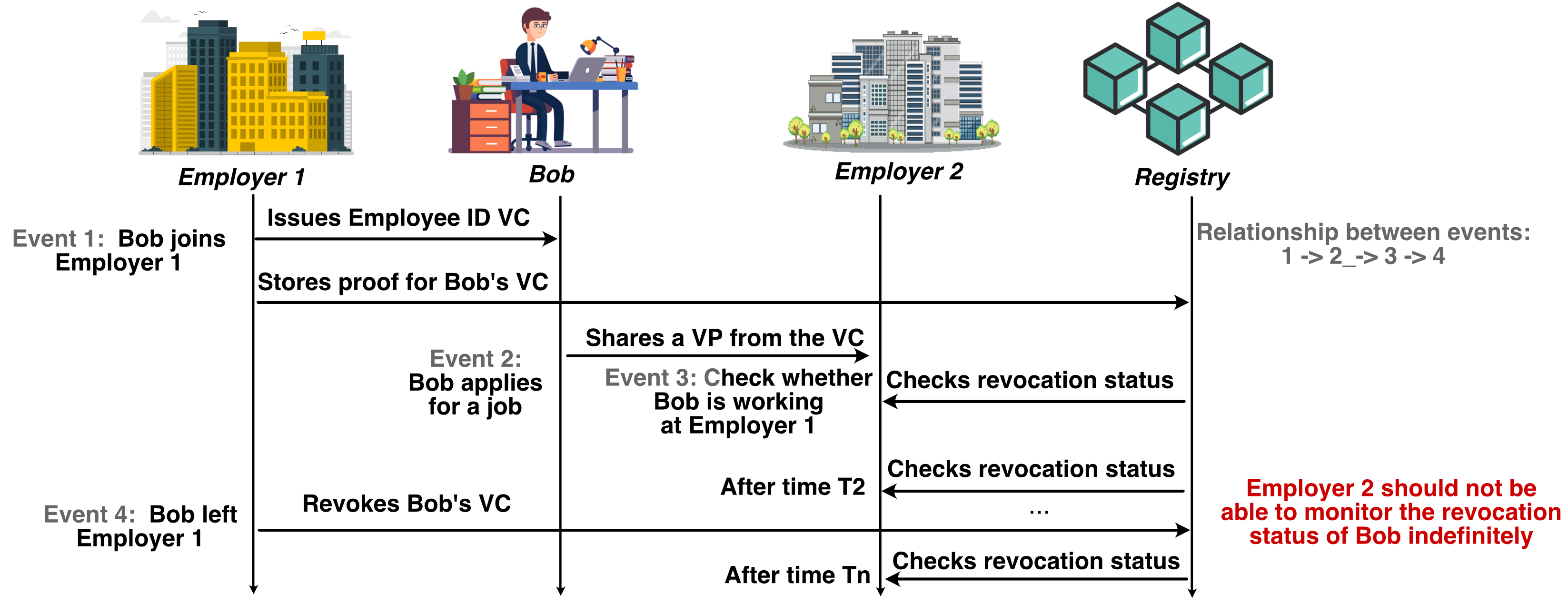}
    \caption{The untraceability requirement}
    \label{fig:problem}
\end{figure*}

Workforce Identity~\cite{credivera,entra, velocity} is one of the popular use cases of \VCs. In Workforce Identity, \VCs are used for the purpose of proving employment status.
Figure~\ref{fig:problem} illustrates a scenario where Employer 1 issues a VC signifying employee ID to Bob who is employed by Employer 1. 
After some time, when Bob applies for a job at Employer 2, Bob generates a \VP from the \VC and shares the \VP with Employer 2 to prove that he is currently working at Employer 1. 
When Bob leaves Employer 1, Employer 1 revokes Bob's Employee ID \VC and stores a proof in the registry.
We refer to and extend this use case throughout this work to motivate the problem and other functional requirements.

\paragraph*{Operations on \VCs.} A system for managing \VCs consists of the following main operations:
\begin{itemize}[noitemsep, topsep=0pt]
    \item \textbf{Issuance} An issuer performs this operation to issue VCs to a holder based on the claims (attributes) provided by the holder.
    \item \textbf{Revocation} An issuer performs this operation to revoke a VC.
    \item \textbf{Presentation} A holder executes this operation to generate a VP for the underlying VC, in order to share the generated VP with a verifier.
    \item \textbf{Verification} A verifier performs this operation to check the validity of the VP received from the holder.
\end{itemize}

VCs may include an expiration period, as specified in the \vcspec~\cite{sporny2021w3cvc}. 
However, expiration is optional and depends on the use case. For example, passports usually expire after a certain period, while diplomas often remain valid indefinitely.

\section{Time-limited continuous verification}
\label{sec:problem}

In existing systems, verifiers use \VPs received from the holders to ascertain the status of underlying \VCs. 
For example, in the Workforce Identity~\cite{credivera,entra, velocity} use case, VCs are used to prove employment status (refer to Section~\ref{sec:background_vc}). 
when Bob applies for a job, he shares a VP with a potential employer. 
The latter verifies the revocation status and expiration period to ascertain the validity of Bob's Workforce Identity. 

The main issue in this scenario is that the verfication can be performed at any point in time. 
This is especially problematic because the verification does not invalidate the \VP; it can be repeated indefinitely until the underlying \VC expires. 
Since many \VCs are issued for months or even years, while certain cerfiticates such as diploma may not expire at all, this allows the verifier to effectively monitor the revocation status of a \VC. 
For example, in the Workforce Identity use case, the potential employer can continuously monitor whether Bob remains employed in his current position.

To address the problem, we introduce the notion of a \verificationperiod for \VPs. 
While a verifier can still perform the verification repeatedly, it can only do it within the \verificationperiod, which is typically much shorter than the expiration period of the underlying \VC. 
The core requirement is that after a \VP expires, the status of the underlying \VC has to be \emph{untraceable} for the verifier.

In order to make time-limited continuous verification practical, the proposed framework includes two additional requirements besides the basic requirement of untraceability:
\begin{asparadesc}
    \item[Configurable verification period:] 
    The holder must be able to configure the \verificationperiod individually for each \VP, even when different \VPs are based on the same underlying \VC. 
    For example, when Bob applies for a job at two different companies, the \verificationperiod might be tailored for each company.
    
    \item[One-time \VP sharing:] 
    Once shared, the \VP contains sufficient information for the verifier to continuously monitor the \VC status during the \verificationperiod without any repeated interaction with the holder. 
    The initial sharing may take multiple communication rounds so that the requirement is different from non-interactive proof. 
    A realistic deployment scale would include thousands of holders and verifiers, making lack of repeated interaction very desirable from the standpoint of bandwidth consumption. 
    
    \item[Unobservable verification:] 
    An issuer should not be able to observe anything about the verification process. 
    This includes the identity of the holder, the identity of the verifier, or even the fact that verification related to some \VC produced by the issuer is taking place. 
    At a basic level, it means that the verifier does not need to contact the issuer for the verification. 
    However, the issuer may potentially be able to learn sensitive information by colluding with the registry, depending on the registry implementation. This is further discussed in Sections~\ref{sec:adversarial} and~\ref{sec:building}.
    
    The requirement of unobservable verification is important for the privacy of any identity management scheme, not just for continuous verification. 
    Note, however, that it is not entirely orthogonal to the proper requirements of time-limited continuous verification. 
    For example, the requirement of one-time \VP sharing may lead to more frequent interactions between verifiers and the registry, thereby imposing further constraints on the registry implementation. 
\end{asparadesc}

It is important to note that the typical challenges of \VC management are further compounded by the above requirements. 
For instance, the requirements imply that the solution would use a variety of cryptographic primitives, yet their overhead must remain practical, even at a large scale of holders and verifiers.

 \subsection{The novelty of the requirements}

Dynamic Status List~\cite{ebsi2024bloomfilter}, Revocation Timeframe 2024~\cite{revocationtimeframe2024}, and Sitouah et al.~\cite{Sitouah2024untrace} consider a limited version of the untraceability problem, with the idea of restricting the verification to be performed within a short period of time after the generation of VPs. 
These works do not provide any formal definitions, though. Besides, they do not consider the requirements of configurable period or single interaction. 
The benefits of these two requirements are evaluated against related work experimentally in Section~\ref{sec:evaluation} and shown to be significant.

The requirement of \textit{unobervable verification} is novel, yet related to the more limited phone-home problem informally mentioned in~\cite{praveen2024prevoke, sporny2021w3cvc, ebsi2023revocationwhitepaper}. The phone-home problem occurs when the verifier directly contacts the issuer in order to verify the VP. In contrast, unobservable verification is a ZK requirement stipulating that the issuer cannot learn anything about the verification.

\section{The \name system}
\label{sec:construction}

The goal of \name is to address the requirements presented in Section~\ref{sec:problem}.
We follow the blacklist approach, storing \tokens corresponding to revoked \VCs in a public ledger. 
The main motivation is providing configurable verification period. However, the blacklist approach also achieves lower bandwidth consumption
because the revocation rate tends to be smaller than the total number of issued \VCs.
In most domains, the revocation rate is very low.
For example, in the Web’s PKI, only $1$\% of certificates are revoked~\cite{schanck2025clubcards, liu2015revocation}. 
Section~\ref{sec:evaluation} discusses the performance aspects in detail. 

In our construction, time is divided into a sequence of \epochs.
All time windows such as the expiration and verification periods are expressed in the number of \epochs. 
The verification period is set by the holder for a \VP of an underlying \VC.

Each entity performs periodic maintenance at the start of every \epoch.
The duration of \epoch is configured as a global system parameter. 
In practice, it should be in the order of minutes or hours: significantly longer than communication delays and sufficiently long so as to amortize the cost of periodic maintenance, yet not too long in order to allow for shorter verification periods.

\subsection{Scheme}
\label{sec:scheme}

\begin{table}[t!]
\caption{Notations}
\label{tab:notations}
\begin{center}
\begin{tabular}{m{1.2cm}m{6cm}}
\toprule
\textbf{Symbol} & \textbf{Description} \\
\midrule
\vc & a VC \\
\midrule
\vp & a VP \\
\midrule
\m & verification period of a VP \\ 
\midrule
\RevList & a list of \seed values corresponding to revoked VCs managed by an issuer \emph{iss} \\
\midrule
\GS & a list of \tokens corr. to revoked VCs \\
\midrule
\tsInit & intial timestamp \\ %
\midrule
\ecounter & current epoch's counter \\
\midrule
\ecounteri{i} & the $i$'th epoch's counter \\
\midrule
\tokenvari{i} & a token corresponding to epoch $i$ \\
\midrule
\durationvar & the duration of a single epoch \\
\midrule
\w & private inputs in the ZK circuit \\
\midrule
\x & public inputs in the ZK circuit \\
\midrule
$\pi$ & a ZK proof \\
\midrule
\crs & a common reference string of a ZK circuit \\
\bottomrule
\end{tabular}
\end{center}
\end{table}

\begin{definition}[\name Scheme]
The \name scheme is a tuple of algorithms: $\Pi$ = (\Setup, \Issuance, \Revocation, \Refresh, \Presentation, \Verification):
\end{definition}
\begin{itemize}[noitemsep, topsep=0pt]
    \item $\mathsf{\Setup}$\,($1^\lambda$) $\rightarrow$ (\sk, \pk, \crs, \tsInit, \durationvar, \RevList, \GS ):
    Given a security parameter, generate and output (a) a secret parameter  \sk, (b) global public parameters  \pk and \crs, (c) the initial timestamp \tsInit, (d) \epoch duration \durationvar (e) an empty list \RevList of revoked VCs, and (f) global state \GS. 
    \item $\mathsf{\Issuance}$\,(\sk, \claims)  $\rightarrow$ \vc: Given (a) a secret parameter \sk, and (b) a set \claims of claims, output \vc, a new VC.
    \item $\mathsf{\Revocation}$\,(\vc, \RevList) $\rightarrow$ \RevListNew: Given \vc and \RevList, output \RevListNew that appends the \seed value encoded in \vc to the \RevList.
    \item $\mathsf{\Refresh}$\,(\ecounter, \RevList) $\rightarrow$ \GS: Given (a) the current \epoch counter value \ecounter, and (b) \RevList, output a list \GS of \tokens corresponding to \seed values in \RevList. 
    \item $\mathsf{\Presentation}$\,(\pk, \crs, \ecounter, \vc, \m, \challenge) $\rightarrow$ \vp: Given (a) \pk and \crs, (b) \ecounter, (c) \vc, (d) \verificationperiod \m in number of \epochs, and (e) \challenge, which is a random number $\in \{0,1\}^*$, output \vp. 
   \item $\mathsf{\Verification}$\,(\pk, \crs, \ecounter, \GS, \m, \vp, \challenge) $\rightarrow \{0, 1 \}$: Given these parameters, output $1$ if \vp is valid in \ecounter and the corresponding VC is neither revoked nor expired. The \vp is valid in \ecounter if it is correctly formed and \ecounter is within the \verificationperiod. Otherwise output $0$.
\end{itemize}

\name extends the standard operations on VCs in the following ways: First, \name includes parameters that affect the validity period of \VPs. Second, \name introduces the \Refresh operation, which is used to support time-limited verification. 
Note also that, in \name, the \Verification operation may be called in a different epoch \ecounter than the epoch $\mathsf{e'}$ in which the \VP was generated. 
The verifier may even decide to perform \Verification repeatedly in different epochs within the \verificationperiod of the \VP.

The implementation presented in Section~\ref{sec:algorithms} does not include details of advanced functionality such as selective disclosure, as the primary focus is on untraceability-related features. Instead, selective disclosure is presented as an extension in Section~\ref{sec:extension_selective_disclosure}.

Table~\ref{tab:notations} lists down notation used throughout the paper.

\subsection{Formal requirements}
A VC scheme should satisfy the three properties of Completness, Soundness, and Untraceability defined as follows:

\begin{definition}[\textbf{Completeness}]
\label{def:completeness}
     Given a \VP valid in epoch \ecounter generated from a non-revoked \VC that is not expired in epoch \ecounter, the verification of \VP in epoch \ecounter should result in success. 
 \end{definition}
Let $\mathsf{(sk, pk, crs, ts_0, dur, RevList_{iss}, blacklist_{iss})} \leftarrow $ $ \mathsf{Setup}(1^\lambda)$.
Let $\mathsf{claims}$ be any valid set of claims, \ecounter be any valid epoch, and $\mathsf{challenge}$ be any valid challenge.
Let $\mathsf{e}'$ and \m be such that $\mathsf{e'} \leq \mathsf{e} \leq (\mathsf{e'}+\mathsf{m})$. 
Let \GSNew be the set of revoked tokens at \ecounter.
Let $C_e$ be the set of non-expired and non-revoked VCs at \ecounter.
A VC scheme satisfies Completeness if the following equality holds for any \vc $\in$ $C_e$, and a VP generated from \vc as \vp $ \leftarrow \mathsf{\Presentation(pk, crs, e', \mathrm{vc}, m, challenge)}$: 
\begin{equation*}
\begin{split}
  & \Pr[\mathsf{\Verification (pk, crs, e, blacklist'_{iss} , m, \mathrm{vp}, challenge)} = 1] \\ & \geq 1 - \mathsf{negl}(\lambda)
\end{split}
\end{equation*}

\begin{definition}[\textbf{Soundness}]
\label{def:soundness}
    Given a VP which is either not valid in epoch \ecounter or created using a revoked or expired VC in epoch \ecounter, the verification of VP in epoch \ecounter should returns $0$.
 \end{definition}
 
 Let $\mathsf{(sk, pk, crs, ts_0, dur, RevList_{iss}, blacklist_{iss})} \leftarrow $ $ \mathsf{Setup}(1^\lambda)$.
Let $\mathsf{claims}$ be any valid set of claims, \ecounter be any valid epoch, and $\mathsf{challenge}$ be any valid challenge.
Let $\mathsf{e'}$ and \m be such that $\mathsf{e'} \leq \mathsf{e} \leq (\mathsf{e'}+\mathsf{m})$. 
Let $C_e$ be the set of non-expired and non-revoked VCs at \ecounter.
Let \GSNew be the set of revoked tokens at \ecounter.
Let $V_e$ be the set of VPs valid in epoch \ecounter generated from VCs in $C_e$ with the use of $\mathsf{pk, crs, m, challenge}$.
A VC scheme satisfies Soundness if for any \vp $ \not \in V_e$ or \vp $\leftarrow$ $\mathsf{\Presentation(pk, crs, e', \mathrm{vc}, m, challenge)}$ where \vc $\not \in C_e$, the following equality holds: 
\begin{equation*}
\begin{split}
  & \Pr[\mathsf{\Verification (pk, crs, e, blacklist'_{iss} , m, \mathrm{vp}, challenge)} = 0]
  \\ & \geq 1 - \mathsf{negl}(\lambda)
\end{split}
\end{equation*}

\begin{definition}[\textbf{Untraceability}]
\label{def:untraceability}
Given a \VP with \verificationperiod $m$,  the verifier should not be able to learn the revocation status of \VC corresponding to the \VP once $m$ is over. 

We formalize untraceability as a game where an adversary $\mathcal{A}$ tries to guess the revocation status of a \VC.
The advantage of $\mathcal{A}$ in the game ($\mathbf{Game}^{untrace}_\mathcal{A}$) (refer to Appendix~\ref{sec: untraceability_security_game}) should be negligible:
\begin{displaymath}
| \Pr[\mathbf{Game}^{untrace}_\mathcal{A}\, (1^\lambda) = 1 ] - 1/2 | \leq \mathsf{negl}(\lambda)
\end{displaymath}
\end{definition}

\subsection{Adversarial model}
\label{sec:adversarial}

We assume that issuers generate \VCs according to the protocol, but may attempt to glean information about the verification process as an attack on \emph{unobservable verification}. 
In other words, issuers are honest but curious. They do not leak information about holders to verifiers.

Similarly, the registry follows the protocol. 
For the sake of \emph{unobservable verification}, we assume that the information stored in the registry is public and read accesses to the registry are not authenticated. Furthermore, the registry should not be able to discern which \VC is being verified. We briefly describe the implementation of such registry in Section~\ref{sec:building}.

We assume that holders follow the protocol with the exception that they may generate fake VPs and proofs during presentation in order to pass the verification by verifiers. For example, the fake VPs may be based on expired, revoked or non-existent \VCs.

Regarding verifiers, we assume that they are trusted to perform the verification operation correctly, but not to preserve the privacy of holders. 
Not only can they attempt to passively extract information from the \VPs and the registry, but they can also actively perform a replay attack. 
Furthermore, a verifier A can collude with a verifier B to learn additional information about holders not available in the \VPs shared with A or B.

To summarize, we consider the entities follow the protocol with the following exceptions: (a) an issuer may try to break \emph{unobservable verification}, (b) a holder might share fake proofs, (c) a verifier may violate the privacy of holders, and (d) a verifier might also impersonate holders. 
This model is similar to, but less benign than the one in \vcspec~\cite{sporny2021w3cvc}, which considers registry and issuers to be fully trusted entities.

\subsection{Additional assumptions}

Expiration periods for \VCs are set by issuers, whereas verification periods for \VPs are set by holders. 
Since verification is performed by verifiers, this requires clock syncronization. Every entity in the system maintains an \epoch counter, which is incremented when the entity transitions to the next \epoch. 
In practice, every entity synchronizes its clock with synchronization servers such as NTP. 
This way, deviations are within seconds, and are thus much smaller than the \epoch duration. We assume that permanent clock failures are detected and repaired. 
However, transient lack of synchronization or communication delays may result in a discrepancy of one \epoch between the counters maintained by different entities, leading to a failed verification for a valid \VP. 
The protocols presented in our solution do not explicitly show how to handle such a situation for the sake of more focused presentation. 
Note that it is easy for holders to overcome such scenarios by creating a new \VP and sharing it with the verifier again, because the clocks and \epoch counters are likely to get synchronized in the meantime.

For the sake of simplified presentation, we assume an issuer uses only one pair (\pk, \sk) of public and secret keys  for signing \VCs, in line with how it is done in the state-of-the-art~\cite{carlo2024evoke, praveen2024prevoke, maram2021candid}. 
It is trivial to extend the presented protocol to support multiple key pairs. 
For example, the public key can be encoded in the \VC and shared in the \VP. 
The set of public keys would be published in the registry.

\subsection{Building Blocks}
\label{sec:building}
Before presenting our \name construction, we describe the buildings blocks of \textbf{\tokens} and \textbf{ZK circuit} used as integral part of the system. 

\paragraph*{Tokens:} 
The system produces \tokens to identify a \VC within a given epoch.
The \tokens are refreshed at the beginning of each \epoch.
A token for an \epoch with counter \ecounter is produced by computing a hash digest \tokenvar $\gets \mathsf{H}$ (\seed, \ecounter) of two inputs, where \seed is a secret value assigned to each \VC by the issuer.
This way, with knowledge of the secret \seed, it is easy to compute future \tokens. 
If a verifier knows a \VC token for a specific \epoch, it will not be able to calculate the next \epoch \token corresponding to the same \VC, which lays the foundation for \emph{untraceability}.
Collision resistance of the hash function ensures uniqueness of \tokens within one \epoch.
A holder will be able to prove validity and ownership of a \token by employing a ZK circuit that proves knowledge of the seed without disclosing it.

\paragraph{ZK circuit:}
\label{sec:circuit}
The specific ZK circuit we use in our construction allows each holder to prove the correctness of his/her \tokens without revealing the \seed. 
The private and public inputs to our ZK circuit are presented in Table~\ref{tab:circuit}. 

\begin{table}[t!]
\caption{ZK circuit inputs}
\label{tab:circuit}
\begin{center}
\begin{tabular}{m{2.7cm}m{4.3cm}}
\toprule
\textbf{Private inputs} (\w) &  \sig,  \seed, \nonce \\
\toprule
 \textbf{Public inputs} (\x) & \pk, \h, \challenge, {\ecounter}, {\validuntil}, {\tokenvar}, $\mathsf{H}\!$ (\claims)
\\
\bottomrule
\end{tabular}
\end{center}
\end{table}

The ZK circuit asserts the following three conditions: 
\begin{enumerate}[noitemsep, topsep=0pt]
    \item $\mathsf{Verify (pk, sig, H(seed, H(claims), exp))} = 1$
    \item $\mathsf{H (seed, e) = token}$
    \item $\mathsf{H (challenge, nonce) = h}$
\end{enumerate}

The first condition verifies the signature (\sig) on (a) \seed, (b) $\mathsf{H}\!$ (\claims), and (c) \validuntil.
These fields are included in a VC.
Since \sig and \seed are private inputs in the ZK circuit, they are never shared with verifiers.
Therefore, this condition enables holders to prove correctness of signature without revealing the signature or any of the underlying fields. 
Thus, this condition binds the tokens with claims.

The second condition verifies that the token is computed correctly.
Since \seed is a private input, holders prove correctness of the \token without revealing the \seed.

The third condition is used to bind each \VP to both the holder and the verifier. The value for \challenge is provided by the verifier, whereas the holder generates a random \nonce and computes a hash digest. 
If this condition is not included, then a malicious verifier could reuse a \VP to impersonate the holder elsewhere, e.g., by presenting it to another service 
(\textit{replay attack}). 
To prevent this, the VP must be bound to both the holder and the verifier.

Our \name construction uses ZK circuit to create a ZK proof for tokens during the creation of a \VP by a holder. Later, the verifier uses the ZK circuit to verify the ZK proof.

\paragraph{Registry:}
The registry is implemented via a DLT, as in~\cite{anoncreds2024} or~\cite{iotaid2025}. Since this is a standard off-the-shelf component, we only describe what information is stored in the registry at what granularity and we elaborate on the interactions between the registry and the other entities, i.e., read and write accesses. 
In particular, it is used to store a record for every issuer that can be retrieved using an issuer's index, e.g., the public key.
To this end, the registry allows for authenticated writes. 
In addition, the registry provides public unauthenticated read access, as mentioned in Section~\ref{sec:adversarial}.
Furthermore, the DLT nodes and other entities should not be able to monitor what data a verifier reads from the registry.
This can be achieved, for example, when the verifier runs its own DLT node.

\subsection{Algorithms}
\label{sec:algorithms}

There are six algorithms in \name: 1) \setup, 2) \issuance, 3) \revocation, 4) \refresh, 5) \presentation, and 6) \verification. 
The pseudo-code is presented in Algorithms~\ref{alg:solution_setup} to~\ref{alg:solution_verification}.

\paragraph{\Setup.}

\begin{algorithm}[t]

\caption{\name: \Setup}
\small
\label{alg:solution_setup}

\begin{algorithmic}[1]
\Procedure {Setup\,}{$1^\lambda$}

\State \label{a:setup-l:time} \tsInit $\leftarrow$ \emph{time}() \Comment{initial timestamp}
\State \label{a:setup-l:epoch} choose \epoch \duration (\durationvar)
\State \label{a:setup-l:keys} (\sk, \pk) $\leftarrow \mathsf{KeyGen}\,(1^{\lambda})$
\State \label{a:setup-l:circuit}Let \emph{circuit} be a ZK circuit asserting: 
\State \indent  $\mathsf{Verify (pk, sig, H(seed, H(claims), exp))} = 1 \ \land$
  \State  \indent  $\mathsf{H (seed, e) = token} \ \land $
   \State \indent $\mathsf{H (challenge, nonce) = h}$ 
\State \label{a:setup-l:crs}\crs  $\leftarrow$ $\mathsf{G16.Setup}$\,($1^{\lambda}$, \emph{circuit})
\State \label{a:setup-l:revlist}\RevList $\leftarrow$ $\phi$ \Comment{stores revoked VCs}
\State \label{a:setup-l:gs}\GS $\leftarrow$ $\phi$
\State output  (\sk, \pk, \crs, \tsInit, \durationvar, \RevList, \GS)

\EndProcedure
\end{algorithmic}
\end{algorithm}

This procedure is used by each issuer to bootstrap the required structures. 
The input $\lambda$ is the security parameter in our protocol. 

First, in L.\ref{a:setup-l:time}-\ref{a:setup-l:epoch}, the issuer records the initial timestemp \tsInit and initiates \epoch duration \durationvar. 
The issuer also generates a cryptographic key pair (\sk, \pk) for the digital signature scheme to be used in the ZK circuit (L.\ref{a:setup-l:keys}).

Subsequently, the issuer defines a ZK circuit to verify the ownership of \tokens and computes a common reference string \crs for the circuit using Groth16~\cite{groth2016zkp} ZKP scheme (L.\ref{a:setup-l:circuit}-\ref{a:setup-l:crs}). 
Then, the issuer initializes two empty lists, \RevList to locally store revoked VCs, and \GS, the global blacklist (L.\ref{a:setup-l:revlist}-\ref{a:setup-l:gs}).

All chosen parameters, keys, and data structures are stored by the issuer.
\GS, \crs, \tsInit, \durationvar, and \pk are published to the registry.
These values are used by verifiers for verification of VPs.

\paragraph{\Issuance.}

\begin{algorithm}[t]

\caption{\name: \Issuance}
\small
\label{alg:solution_issuance}

\begin{algorithmic}[1]
\Procedure {Issuance\,}{\sk, \claims}

\State \seed $\leftarrow$ $\{ 0,1\}^\lambda$ \Comment{cryptographically secure value}
\State choose \validuntil

\State \sig $\leftarrow$ $\mathsf{Sign}\,$(\sk,  $\mathsf{H}\!$  (\seed, $\mathsf{H}\!$ (\claims), \validuntil))

\State \vc $\leftarrow$ (\seed, \claims, \validuntil,  \sig)
\State  output \vc
\EndProcedure
\end{algorithmic}
\end{algorithm}

This procedure is used by an issuer to generate a new VC for a holder. 
It takes the issuer's secret key \sk and a set \claims of claims  as inputs.
Depending on the use case, the set of claims can be provided either by the issuer or by the holder or collectively by both of them.

To issue a new VC, first, the issuer samples a $\lambda$-bit \seed uniformly. 
Then, the issuer determines the \expirationperiod denoted as \validuntil for the new VC. 
The issuer encodes these values in the VC. In addition, the issuer can include a unique identifier \emph{id} for the VC. 
We do not mandate including \emph{id} in a VC since this field is optional in the \vcspec.

Signing raw data would increase the size of the ZK circuit leading to reduced performance.
Therefore, the signature is performed on a hash digest to keep the size of the messages fixed.
The issuer computes a hash digest of the following data fields and generates a digital signature (\sig): (a)  \seed, (b) $\mathsf{H}\!$ (\claims), and (c) \validuntil. 
Finally, a new VC \vc is created; it consists of the signed data fields along with the signature.
Finally, the procedure outputs the \vc.

Once the new VC \vc is generated, the issuer sends \vc to the corresponding holder along with \crs, \pk, \tsInit, and \durationvar. 
The holder will use these values to generate proofs while creating a \VP (Algorithm~\ref{alg:solution_presentation}).  

\paragraph{\Revocation.}

\begin{algorithm}[t]

\caption{\name: \Revocation}
\small
\label{alg:solution_revocation}

\begin{algorithmic}[1]
\Procedure {Revocation\,}{\vc, \RevList}

\State \RevListNew $\leftarrow$ \RevList $\cup$ \vc.\seed
\State output \RevListNew
\EndProcedure
\end{algorithmic}
\end{algorithm}

This procedure is used by an issuer to revoke a previously issued VC. 
When revoking a given VC, the issuer appends the corresponding \seed to \RevList, a list that stores the seeds of revoked VCs.
The procedure then outputs the updated list \RevListNew.
The issuer stores the updated list locally.

While revocation is most commonly used when the holder's status changes, there could be additional technical reasons.
In practice, it is essential to revoke a VC if the underlying \seed is leaked because adversaries may impersonate the holder by generating ZK proofs using the \seed.

\paragraph{\Refresh.}

\begin{algorithm}[t]

\caption{\name: \Refresh}
\small
\label{alg:solution_refresh}

\begin{algorithmic}[1]
\Procedure {Refresh\,}{\ecounter, \RevList}

\State \GS $\leftarrow$ $\phi$ 
\For{\vc $\in$ \RevList}
\State \tokenvar $\leftarrow  \mathsf{H}\!$ (\vc.\seed, \ecounter) \Comment{updates global state}
\State \GS $\leftarrow$ \GS $\cup$ \tokenvar
\EndFor
\State output \GS

\EndProcedure
\end{algorithmic}
\end{algorithm}

This procedure is used by an issuer to periodically refresh \tokens corresponding to revoked VCs. 
It is executed every time a new \epoch starts. 
At the start of a new \epoch, assuming the local time of $\mathsf{current\_time}$, the issuer first computes the current \epoch value (\ecounter) as follows: 
\begin{equation}
\mathsf{e} \leftarrow (\mathsf{current\_time} - \mathsf{ts}_0)/\mathsf{dur}
\label{formula:current_epoch}
\end{equation}
Then, the issuer uses the \Refresh procedure to compute \tokens corresponding to revoked VCs.
After that, the issuer sends the \tokens to the registry, updating the global blacklist \GS. 

Before starting the procedure, the issuer is expected to clean up seeds corresponding to expired VCs from \RevList. 

We note that revocation of VCs take effect only after a \Refresh operation is performed, since during \Refresh, tokens corresponding to seeds added to \RevList are published in the \GS.
The issuer must perform \Refresh at the start of each epoch.

\paragraph{\Presentation.}

A holder uses this procedure to generate a \VP that will be sent to a verifier. 
The procedure takes the following inputs: (a) \pk, (b) \crs, (c) \ecounter, (d) \vc, (e) \m, and (f) \challenge. 
Parameter \vc denotes the holder's \VC issued by the issuer. The holder receives \pk and \crs from the issuer alongside the \vc.

Parameter \m specifies the \verificationperiod of the new \VP \vp in terms of the number of \epochs starting from the current \ecounter. It translates into the number of tokens that the holder wants to include in \vp.

Before generating \vp, the holder receives \challenge from the verifier. 
The parameter \challenge is used to prevent replay attacks on the proofs in \vp.

\begin{algorithm}[t]

\caption{\name: \Presentation}
\small
\label{alg:solution_presentation}

\begin{algorithmic}[1]
\Procedure {Presentation\,}{\pk, \crs,  \ecounter,  \vc, \m, \challenge}
\State \nonce $\leftarrow$ $\{0,1\}^\lambda$
\State \h $\leftarrow$ $\mathsf{H}\!$ (\challenge, \nonce)
\State \ecounters $\leftarrow$ $\phi$ 
\State \tokenvars  $\leftarrow$ $\phi$ 
\State $\pi_\mathsf{zkp}$  $\leftarrow$ $\phi$ 

\For{$i \in 1, \cdots,$ \m}
\State \ecounteri{i}  $\leftarrow$ \ecounter 
\State \tokenvari{i} $\leftarrow \mathsf{H}\!$ (\vc.\seed, \ecounteri{i})
\State \w $\leftarrow$ (\vc.\seed, \vc.\sig, \nonce)
\State \x $\leftarrow$ (\pk, \tokenvari{i}, \ecounteri{i}, \challenge, \h, \vc.\validuntil, \vc.\claims)
\State $\pi_i \leftarrow \mathsf{G16.Prove}$\,(\x, \crs, \w)  
\State  $\pi_\mathsf{zkp}$  $\leftarrow$ $\pi_\mathsf{zkp}$ $\cup$ $\pi_i$
\State \ecounters $\leftarrow$ \ecounters $\cup$ \ecounteri{i}
\State \tokenvars $\leftarrow$  \tokenvars $\cup$ \tokenvari{i}
\State \ecounter  $\leftarrow$ \ecounter  $+ 1$
\EndFor

\State \vp $\leftarrow$ (\tokenvars, \ecounters, \m, \h, \vc.\claims, \vc.\validuntil, $\pi_\mathsf{zkp}$) 
\State output \vp 
\EndProcedure
\end{algorithmic}
\end{algorithm}

The Presentation procedure is shown in Algorithm~\ref{alg:solution_presentation}.
First, the holder computes the current epoch value using Equation~\ref{formula:current_epoch}. 
Subsequently, the holder generates a random \nonce. Then, it computes a hash digest \h of two inputs: \challenge provided by the verifier and \nonce.

The holder generates \m tokens, which act as pseudonyms representing the holder for \m consecutive \epochs starting from the current one. 
Since \ecounter acts as a counter, the holder can simply increment \ecounter by $1$ to compute the next \epoch index. 
In addition to \tokens, the holder generates ZK Proofs $\pi_\mathsf{zkp}$ to prove the correctness of created \tokens. 
For each \tokenvari{i}, the holder generates a Zero-Knowledge Proof $\pi_i$ using private and public inputs.
The private (\w) and public (\x) inputs are described in Table~\ref{tab:circuit}.
$\pi_i$ proves that \say{The VC encodes a signed \seed using which the provided \tokenvari{i} is generated.}
We use $\pi_i$ to bind \vp to both the holder and the verifier, see the third condition for the ZK circuit in Section~\ref{sec:circuit}. 
The proofs are accumulated into $\pi_\mathsf{zkp}$. 
Using ZK Proofs allows the holder to prove the ownership of \tokens without revealing the underlying \seed. 

Finally, the holder constructs \vp consisting of: (a) \m  \tokenvars, (b) \m \ecounters, (c) \m, (d) \h, (e) \vc.\claims, (f) \vc.\validuntil, and (g) $\pi_\mathsf{zkp}$. 
After \vp is generated, the holder may share it with a verifier. 
The holder also shares identifying information of the issuer that allows the verifier to extract the issuer's record from the registry.

\paragraph{\Verification.}

\begin{algorithm}[t]

\caption{\name: \Verification}
\small
\label{alg:solution_verification}

\begin{algorithmic}[1]

\Procedure {Verification\,}{\pk, \crs, \ecounter, \GS, \m, \vp, \challenge}
  \If{ \textsc{ \footnotesize VerifyProofs}(\pk, \crs, \m, \vp, \challenge) $\neq$ $0$}
    \State output \textsc{ \footnotesize VerifyRevocationStatus}(\ecounter, \vp, \GS)
  \EndIf
  \State output $0$
\EndProcedure
\newline 

\Procedure {VerifyProofs\,}{\pk, \crs, \m, \vp, \challenge}

\For{$i \in 1, \cdots,$ \m}
\State \ecounteri{i} $\leftarrow$ extract the counter from \vp.\ecounters corr. to $i$ 
  \State \tokenvari{i} $\leftarrow$ extract the \token from \vp.\tokenvars  corr. to $i$ 
    \State $\pi_i \leftarrow$ extract the $\pi$ from \vp.$\pi_\mathsf{zkp}$  corr. to $i$ 
  \State \x $\leftarrow$ (\pk, \tokenvari{i}, \ecounteri{i}, \challenge, \vp.\h, \vp.\vc.\validuntil,  $\mathsf{H}\!$ (\vp.\vc.\claims))
  \If{ $\mathsf{G16.Verify}$\,(\x, \crs, $\pi_i$) $= 0$}
    \State output $0$
  \EndIf
  
\EndFor
\State output $1$

\EndProcedure
\newline 
\Procedure {VerifyRevocationStatus\,}{\ecounter, \vp, \GS}

\If{\ecounter $\notin$ \vp.\ecounters $\wedge$ \ecounter > \vp.\vc.\validuntil}
  \State output $0$
\EndIf
\State \tokenvar $\leftarrow$ extract the \token from \vp.\tokenvars  corr. to \ecounter
\If{\tokenvar $\in$ \GS}
  \State output $0$
\EndIf
\State output $1$
\EndProcedure

\end{algorithmic}
\end{algorithm}

The input parameter \challenge to the \verification procedure should have the same value that was previously sent by the verifier to the holder.
First, the verifier retrieves the issuer's record from the registry including the public key \pk, the common reference string \crs, and the global blacklist \GS.
Then, the verifier computes the current epoch value using Equation~\ref{formula:current_epoch}.

Afterwards, the verifier performs the steps shown in Algorithm~\ref{alg:solution_verification}.  
\textit{Verification} includes two procedures: \textit{VerifyProofs} and \textit{VerifyRevocationStatus}.
\textit{VerifyProofs} checks the correctness of proofs included in \vp.
The proofs are verified for all epochs included in \vp. For each proof, the verification checks that  the claims and the secret seed have been signed by the issuer, and that the \token is computed correctly. The procedure returns $1$ if all proofs are valid, and $0$ otherwise.
\textit{VerifyProofs} checks all proofs included in \vp, not just the proof belonging to the current epoch, to ensure they are not malformed.

In contrast, \textit{VerifyRevocationStatus} checks the revocation status of \vp only in the current epoch using the current blacklist \GS retrieved from the registry.
The procedure also checks that \vc underlying shared \vp is not expired.

To explain the assymetry between \textit{VerifyProofs} and \textit{VerifyRevocationStatus}, recall the verification semantics for VPs: the verification fails, if \vp contains a malformed or incorrect proof for any epoch, or if \vp is invalid in the current epoch.
As an optimization, if the verifier decides to call \textit{Verification} on same \vp repeatedly for different epochs, it is sufficient to invoke \textit{VerifyProofs} only the first time, whereas \textit{VerifyRevocationStatus} needs to be called each time using the current values of \ecounter and \GS.

\subsection{Extensions}
\label{sec:extensions}
\label{sec:extension_selective_disclosure}

So far, we have been presented the core functionality of \name that focuses on implementing untraceability. In this section, we provide three extensions to the core functionality: provision for selective trust in issuers, support for selective disclosure of claims, and a single ZK proof covering multiple \tokens.

\paragraph{Selective Trust in Issuers.}

In practice, verifiers have the flexibility to choose which issuers to trust based on their individual credibility~\cite{sporny2021w3cvc, praveen2025survey}. The \textit{Verification} function may reject a VP based on a VC from an untrusted issuer. 
This trust is determined at a policy-level. For example, a potential employer (verifier) might decide to accept workforce identity VCs issued by organizations (issuers) known only to the employer. 
This is orthogonal to the core functionality of time-limited verification, though it requires specific straightforward adjustments to the scheme interface and definitions in Section~\ref{sec:scheme}, as well as to the implementation of \textit{Verification}.

\paragraph{Selective Disclosure.}

Selective disclosure means that only a subset of claims included into a VC is transferred into a VP. 
While not directly related to our objectives, it is one of the key requirements in many VC management systems, as observed in~\cite{seila2024survey}. 
Our approach employs a mechanism similar to the one used in IOTA Identity: Revocation Bitmap 2022~\cite{revocationbitmap2022}. 
Each holder can choose which claims to reveal and which to hide. 
It hides a claim by only revealing the hash digest of that claim. 
For example, if a VC includes a name and a date-of-birth, a VP can include the date-of-birth and the hash digest of the name.

Thus, VPs include hash digest of claims instead of claims themselves.
Similarly, the signature is computed on the hash digest of individual claims rather than on all claims collectively. 
When sharing the digest with a verifier, the holder additionally sends the claims selected for disclosure and the index of those claims in the hash digest.
The verifier can verify the correctness of a claim by computing its hash digest and matching the computed digest against the provided one.

Since digital signatures are verified on the ZK circuit, the verification function is updated to verify the signature of the hash digest of claims instead of the claims themselves. 
The public input of $\mathsf{H}\!$  (claims) should be replaced with the hash digest of individual claims.

\paragraph{Proving Multiple Tokens using a Single ZK Proof.}

In the algorithmic description above, a separate ZK proof is used for each \token. 
Below, we provide an optimization that allows a holder to prove multiple \tokens using a single ZK proof. 
During the \setup procedure, before generating the circuit, an issuer has to choose $k$ that specifies the number of \tokens that can be verified in the ZK circuit.
The issuer needs to store $k$ in the registry so that verifiers can fetch it during \verification.

The ZK circuit should be updated as follows: the condition of  $\mathsf{H (seed, e) = token}$ in the ZK circuit defined in Section~\ref{sec:circuit} is replaced with the following condition: $\mathsf{H}\!$ (\seed, \ecounteri{i}) $=$ \tokenvari{i} $|\, \forall i \in 1 \cdots k$. 
The public input \ecounter in the ZK circuit is replaced with \ecounteri{i} $|\, \forall i \in 1 \cdots k$.
The public input \tokenvar is replaced with  \tokenvari{i} $|\, \forall i \in 1 \cdots k$.

The holder generates \tokens in blocks of \emph{k} because the ZK circuit requires exactly \emph{k} \tokens.
If \m is not a multiple of \emph{k}, then the last \token (\tokenvarm) is duplicated to fill the remaining spaces in the last block. 
For each block of \emph{k} \tokens, the holder generates a Zero-Knowledge Proof $\pi_i$ using private and public inputs.
The latter proves the following statement: {The VC encodes a signed \seed using which the provided \tokenvars are generated}.

However, this optimization has three challenges. 
First, before generating a ZK circuit, the value of \emph{k} needs to be fixed. Therefore, an issuer needs to select a single value of \emph{k} that would be used for all VCs generated by that issuer. 
Second, increasing the value of \emph{k} directly results in increased ZK circuit size, which affects the bandwidth of issuers, holders, and verifiers. 
Third, the circuit generation time would also be affected by increasing the value of \emph{k}. 
Section~\ref{sec:evaluation_zkp} experimentally evaluates the impact of \emph{k} and discusses the results. 

\section{Security Analysis}
\label{sec:theorems}
We prove that \name satisfies completeness, soundness and untraceability properties even in the presence of a computationally-bounded adversary $\mathcal{A}$. 

\begin{theorem}
If the ZKP system is Complete and the hash function is collision-resistant, then $\Pi$ = (\Setup, \Issuance, \Revocation, \Refresh, \Presentation, \Verification) satisfies the Completeness property. 
\end{theorem}

\begin{theorem}
If the ZKP system is sound and Zero-Knowledge, the digital signature scheme is Unforgeable, and the hash function is pre-image resistant, then $\Pi$ = (\Setup, \Issuance, \Revocation, \Refresh, \Presentation, \Verification) satisfies the Soundness property. 
\end{theorem}

\begin{theorem}
If the ZKP system is Zero-Knowledge and the hash function is pre-image resistant, then $\Pi$ = (\Setup, \Issuance, \Revocation, \Refresh, \Presentation, \Verification) satisfies the Untraceability property. 
\end{theorem}

The proofs are described in Appendix~\ref{sec:security}.

\section{Discussion}
\label{sec:discussion}

In this section, we discuss how \name satisfies additional requirements other than the three analyzed in Section~\ref{sec:theorems}.

\paragraph{Configurable Verification Period.}

Before sharing a VP with a verifier, a holder and the verifier determine \m, the \verificationperiod of the VP expressed in the number of epochs.
The holder generates a VP consisting of \m \tokens corresponding to \m \epochs starting from the current one.
Therefore, configurability of \verificationperiod is provided by the construction.

It is also possible for holders to include \tokens for specific \epochs instead of including a sequence of \m \tokens starting from the current \epoch \ecounter. To do this, the \Presentation and \Verification algorithms need to be modified. First, in the \Presentation, the input parameter \m should be replaced with specific \ecounter values. 
Then, the corresponding \tokens should be generated instead of generating a sequence of \m \tokens. Finally, the \VP should include \ecounter values instead of \m. 
In the \Verification procedure, a verifier needs to verify the correctness of the included \ecounter values. 

\paragraph{One-time VP sharing.}

A holder includes one or more ZK proofs ($\pi_\mathsf{zkp}$) in a VP.
$\pi_\mathsf{zkp}$ is sufficient for a verifier to ascertain the validity of \tokens included in the VP.
Therefore, further interactions are not required for continuous verification in our protocol.

\paragraph{Unobservable Verification.}

We provide an informal intuition about how \name achieves unobservable verification.
An issuer publishes the revocation list and other public parameters in the registry, as explained in Section~\ref{sec:building}. 
The verifier retrieves this information from the registry and verifies VPs. 
Registry provides public read access where no access permissions are required to read information. 
In addition, no identifiers for VCs are stored in registry. 
Therefore, an issuer does not learn about the verification process.

\paragraph*{Unlinkability.}

A number of systems~\cite{irma2024revocation,anoncreds2024} and papers~\cite{rosenberg2023zkcreds} have considered an auxiliary design goal of unlinkability. Unlinkability is defined in \vcspec~\cite{vc} as the property  ensuring that colluding verifiers cannot correlate the VPs they receive.

While unlikability is not a design goal of our work, \name can be extended to provide a limited form of unlinkability.
To do that, first, the selective disclosure extension described in Section~\ref{sec:extension_selective_disclosure} should be updated so that no identifying information is shared with verifiers.
In the current extension, hash digests of all claims are shared with verifiers, which could be used to correlate multiple VPs. 
This could be addressed by randomizing the hash digests shared with the verifiers by a nonce, and proving the correctness of these digests in the ZKP.
Once no correlatable information is shared, \name would provide unlinkability when multiple VPs from the same VC do not contain overlapping \verificationperiod. 

It is also possible to provide unlinkability for a bounded number of VPs with overlapping \verificationperiod by assigning multiple \seed values to a VC. 
A holder can generate unlinkable VPs with overlapping \verificationperiod by using different \seed values.
The cost of this approach is that the size of the blacklist would be increased proportional to the number of seeds per VC.

\section{Evaluation}
\label{sec:evaluation}

In this section, we evaluate \name against related work. 
Our main objectives are listed below:
\begin{itemize}[noitemsep, topsep=0pt]
    \item What is the impact of providing configurable period and one-time VP sharing on performance?
    \item What are the performance tradeoffs of following a blacklist approach?
    \item What are the benefits and costs of using general-purpose ZKPs compared to custom-built ones?
\end{itemize}
These objectives reflect the key differences between \name and related work.

We measure bandwidth and computation requirements on different entities for sharing and verifying VPs, and 
for refreshing the blacklist every epoch.
These operations are most relevant to revocation protocols. 
We also evaluate the storage requirements at the registry and the impact of proving multiple tokens using a single ZK proof, as presented in Section~\ref{sec:extensions}.
For each objective, we focus on the metrics and entities that are most impacted.

\subsection{Workload}

We consider the use case of Workforce Identity (refer to Section~\ref{sec:background_vc}). 
We are not aware of any workload analysis for this use case. Therefore, we adopt the scale and parameters from published works on similar use cases, and complement it with sensitivity analysis.

We consider the scale of one million issued VCs in total, similar to EVOKE~\cite{carlo2024evoke}.
The \expirationperiod of VCs is set to $365$ days: We assume that an employer would update the identities of employees after each financial year for reasons such as salary increase, promotion, etc. The duration of \epoch is set to one day. Such refresh duration is common in existing systems, such as Anoncreds. In IRMA~\cite{irma2024revocation}, holders fetch updated witnesses from issuer every day.

Similarly to the evaluation in~EVOKE~\cite{carlo2024evoke} and Sitouah et al~\cite{Sitouah2024untrace}, we only consider revocation of VCs in the workload, without expiration. We evaluate revocation rate $\mathcal{R}$ from $1\%$ to $15\%$.
This range is very close to the one considered by EVOKE~\cite{carlo2024evoke}, which is based on studies that analyze PKI certificate revocation. 
In experiments, we assume a fixed number $r$ of VCs are revoked per epoch.
For example with 1 million VCs, expiration period of $356$ epochs (days), and $\mathcal{R}=1\%$ revocation rate, results in revoking $r\approx 27$  VCs each epoch.

For credentials shared during a hiring process, we assume that verification periods between one day and two months are reasonable, depending on the hiring process.
Thus, we vary the verification period \m from $1$ to $60$ epochs.

When evaluating the extension of using a single ZK proof for \emph{k} tokens, we consider the values of \emph{k} from $1$ to $2^5$, since values of \emph{k} larger than \m create overhead without benefits.

\subsection{Baseline Solution}

We opt for IRMA~\cite{irma2024revocation} as a baseline due to the following reasons:
\begin{asparaitem}
	\item The guarantees listed in Table~\ref{tab:related_work} and Section~\ref{sec:problem} and provided by \name come at a performance cost due to the use of complex techniques. To allow for a fair comparison, a system with similar guarantees needs to be selected as a baseline. Since no state-of-the-art system provides all listed guarantees, we choose IRMA that implements separation separation between expiration and verification periods alongside a form of untraceability.
	\item IRMA is a popular system deployed in real-world use cases. Furthermore, it provides well-documented open-source libraries.
	\item IRMA is a good match for our evaluation objectives. In contrast to \name, it does not have one-time sharing (Objective 1), it is based on a whitelist (Objective 2), and it uses custom-built ZKPs (Objective 3).
\end{asparaitem}

To achieve similar performance configuration in IRMA compared to \name, we let the issuer update the accumulator once every epoch, removing factors from revoked VCs.
In IRMA, when the accumulator is updated, the issuer broadcasts all revoked factors to the holders, including both newly and older revoked factors. 
This allows holders to receive the factors also after offline periods.
To account for different deployments, we also investigate two additional settings, \emph{no repetition} and \emph{registry}.
In IRMA-\emph{no repetition}, every revoked factor is broadcast once to all holders. 
In IRMA-\emph{registry}, instead of broadcasting, revoked factors are published in the registry, making them available to holders.
This variant significantly differs from IRMA, which does not use DLTs, but is similar to approaches in other accumulator-based protocols, such as Anoncreds~\cite{anoncreds2024}, and Sitouah et al~\cite{Sitouah2024untrace}.

\subsection{One-time Sharing}
\label{sec:one_time_sharing}

Our first objective is to measure the benefits of the one-time sharing feature. 
We compare the bandwidth and computation requirements of holders, since the one-time sharing feature requires holders to compute and share tokens and proofs corresponding to \m epochs provided in the \verificationperiod. 
To understand the results, it is important to note that we compare the overhead of sharing a single VP with verification period \m in \name with the overhead of repeatedly sharing proofs in IRMA, during \m consecutive epochs.
In addition to repeatedly computing and sharing the VP in IRMA, the holder also 
needs to receive a witness update from an issuer, and update the proof accordingly.

\subsubsection*{Metric 1: Holder's Bandwidth}

In \name, we measure the total bytes for ZK proofs sent by the holder.
In IRMA, we measure the total bytes of witness update messages a holder receives from an issuer during the validity period, in addition to the bytes for proofs sent to the verifier.

Figure~\ref{fig:result_holder_bandwidth} illustrates the bandwidth requirements for holders. 
In \name and IRMA-no repetition, the bandwidth requirement is linear. 
\name outperforms IRMA by a huge margin.
For example, when \m is configured to $26$ epochs, the required bandwidth in IRMA-\emph{no repetition} is around $85$ KB, whereas it is $4$ KB in \name. 
In IRMA with broadcast everything, the bandwidth is $415$ KB.
The bandwidth is further increased when increasing the revocation rate in IRMA. 
IRMA-\emph{registry} variant is not shown, but values will be the same as the ones used in IRMA-\emph{no repetition}, since a holder can fetch the same amount of proofs directly from the registry instead of receiving it from an issuer.

\subsubsection*{Metric 2: Holder's Computation}

\begin{figure*}[t]
     \centering
\begin{subfigure}[t]{0.33\textwidth}
    \centering
    \includegraphics[height=4.8cm,width=\textwidth]{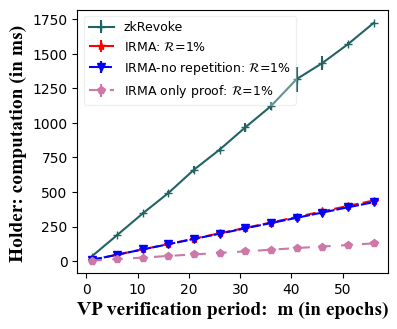}
    \caption{Impact of small $\mathcal{R}$ on computation}
            \label{fig:result_one_time_sharing_computation}
\end{subfigure}
\begin{subfigure}[t]{0.33\textwidth}
    \centering
    \includegraphics[height=4.8cm,width=\textwidth]{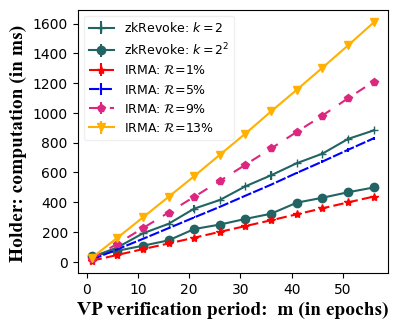}
    \caption{Impact of increasing $\mathcal{R}$, $k$ on computation}
\label{fig:result_one_time_sharing_computation_k2}
\end{subfigure}
\hfill
\begin{subfigure}[t]{0.33\textwidth}
    \centering
    \includegraphics[height=4.8cm,width=\textwidth]{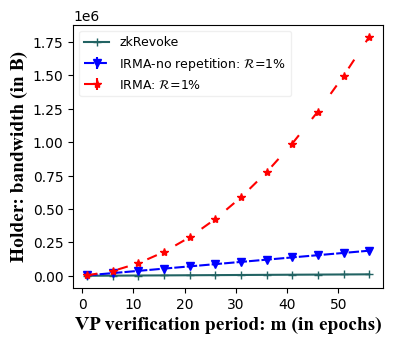}
    \caption{The bandwidth requirements for holders}
\label{fig:result_holder_bandwidth}
\end{subfigure}
\hfill

 \caption{The computation and bandwidth overheads of proof sharing for a single VP}
\end{figure*}

In \name, we measure the total amount of time a holder needs to compute ZK proofs. 
For IRMA, where a holder updates witnesses and recomputes proofs every epoch, we report the total amount of time spend, including witness updates.

Figure~\ref{fig:result_one_time_sharing_computation} shows that IRMA outperforms \name for small revocation rate $\mathcal{R}$. 
For example, consider the case when \m is set to one. In \irma, a holder takes $8$ milliseconds, whereas in \name, it takes $32$ milliseconds so that the computation time in \irma is four times shorter. 
The results for computing only proofs show that approximately 70\% of the computation done in IRMA is due to witness updates.

Higher revocation rates lead to increased computation time in IRMA since the holder needs to incorporate more witness updates.
The revocation rate does not impact the holder computation in \name.
In addition, using the extension of proving $k$ tokens using a single ZK proof reduces the computation time in \name.
Figure~\ref{fig:result_one_time_sharing_computation_k2} shows the impact of increasing $k$ and $\mathcal{R}$.
For example, when $k=4$ and $\mathcal{R}>5$, IRMA requires more computation from the holder.

\paragraph*{Summary.}

Holders in \name require significantly less bandwidth compared to IRMA since constructing VPs in  \name does not depend on the revocations of other VCs, unlike in IRMA. 
Similarly, the computation for holders in \name does not depend on the revocation of other VCs, in contrast to IRMA. 
When the extension of proving multiple tokens is used, \name incurs computation overhead for holders that is comparable to or better than the one in IRMA.

\subsection{Blacklist approach}

\begin{figure*}[t]
     \centering
\begin{subfigure}[t]{0.33\textwidth}
    \centering
    \includegraphics[height=5cm,width=\textwidth]{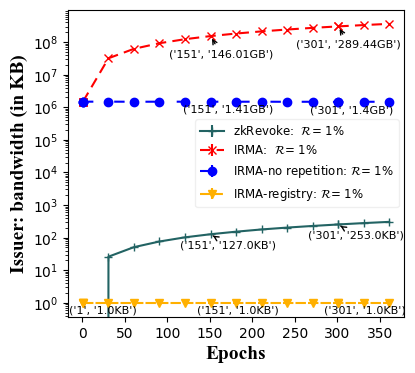}
    \caption{The bandwidth requirements for issuers}
    \label{fig:result_revocation_issuer_bandwidth_all}
\end{subfigure}
\hfill
\begin{subfigure}[t]{0.33\textwidth}
    \centering
    \includegraphics[height=5cm,width=\textwidth]{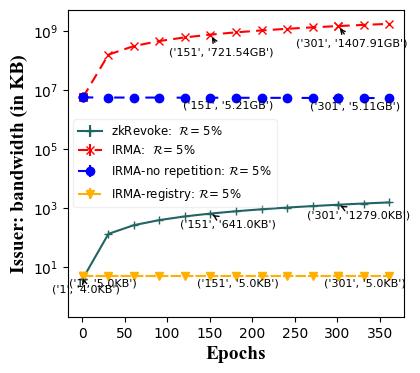}
     \caption{Impact of increasing $\mathcal{R}$, $k$ on bandwidth}
    \label{fig:result_revocation_issuer_bandwidth_r5}
\end{subfigure}
\hfill
\begin{subfigure}[t]{0.33\textwidth}
    \centering
    \includegraphics[height=5cm,width=\textwidth]{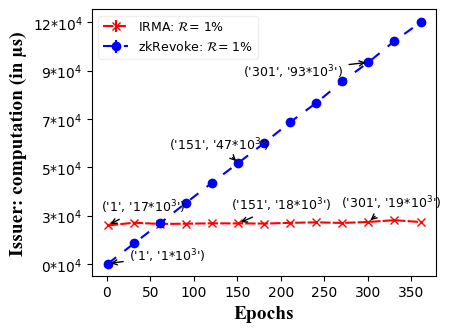}
    \caption{The computations requirements for issuers}
    \label{fig:result_revocation_issuer_computation}
\end{subfigure}

 \caption{The performance impact of blacklist vs. accumulators on revocation}
 \label{fig:blacklist}
\end{figure*}

Our second objective is to quantify the benefits and costs of following a blacklist approach compared to a whitelist.
We compare computation and bandwidth requirements of issuers, since issuers are the ones who update the list and share the list periodically irrespective of whether a blacklist or whitelist is used. 
In addition, since storage in DLTs typically incurs high cost, we also evaluate the storage requirements at the registry.
The overhead for holders was already covered in Section~\ref{sec:one_time_sharing}, while the bandwidth overhead for verifiers to download lists can be inferred from the storage overhead at the registry.

In Figure~\ref{fig:blacklist}, we show how the performance impact on the issuer evolves over $365$ epochs as more VCs are revoked.

 \subsubsection*{Metric 3: Issuer's Bandwidth}
Figure~\ref{fig:result_revocation_issuer_bandwidth_all} shows the bandwidth requirements for issuers. In the experiment, every day $r=27$ VCs are revoked, eventually reaching 10000 revoked VCs ($\mathcal{R}=1\%$).

In \name, the issuer publishes the blacklist every epoch, which grows in size as more VCs are revoked. The blacklist eventually reaches $\mathcal{R} \times n \times 32 \ \textnormal{bytes} $ = $\Theta(\mathcal{R})$.

In IRMA, the issuer publishes the accumulator every epoch, and sends all revoked factors to the holders of valid VCs.
Even without repetition (IRMA-\emph{no repetition}) this requires significant bandwidth $\Theta(r\cdot n)$ from the issuer.

In IRMA-\emph{registry}, the issuer only publishes the new accumulator and the new revoked factors to the registry.
Thus, the issuer bandwidth requirement does not depend on the total number of revoked VCs ($\mathcal{R}$) but only on the number of VCs revoked per \epoch ($r$).

 Figure~\ref{fig:result_revocation_issuer_bandwidth_r5} shows that increased revocation rate leads to higher bandwidth consumption in \name, but the difference from the variants of IRMA stays the same.

\paragraph*{Metric 4: Issuer's Computation.}
In IRMA, we measure the time to update accumulator in each epoch.
In \name, we measure the time to compute tokens each epoch.
Figure~\ref{fig:result_revocation_issuer_computation} plots the results.
The time to perform \refresh operation in \name increases linearly over the period of time since revoked VCs are accumulating in the blacklist.
For example, \name requires more compuation than IRMA by a factor of $3$ in epoch $201$ and by a factor of $5$ in epoch $301$.

\paragraph*{Metric 5: Storage.}
We provide only theoretical analysis for storage requirements since it mostly depends on the number of hash values stored, not on the implementation. 
In IRMA, accumulators require a constant amount $\Theta(1)$ of storage.
In \name, the blacklist requires linear amount $\Theta(\mathcal{R}\cdot n)$ of storage where $\mathcal{R}$ is the fraction of valid VCs which are revoked and not expired during an epoch.
The variant IRMA-\emph{registry} requires the same amount of storage as \name, since the revoked factors are stored in the registry.

\paragraph*{Summary.}

Our evaluation shows that issuers need to perform more computation in \name compared to IRMA, since the issuer has to recompute tokens for all revoked VCs after each \epoch. 
\name avoids significant bandwidth consumption required in IRMA to broadcast factors, but requires more bandwidth than IRMA-\emph{registry} since the blacklist is replaced every epoch, not just appended to.

\begin{figure*}[t]
     \centering
     \begin{subfigure}[t]{0.33\textwidth}
    \centering
    \includegraphics[height=4.8cm,width=\textwidth]{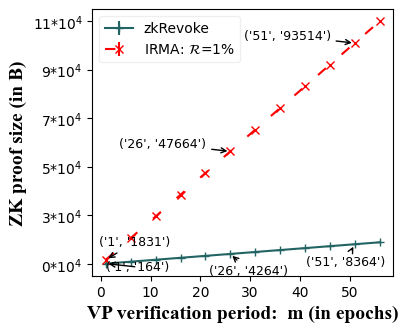}
    \caption{The total amount of proofs in a VP}
    \label{fig:result_zkp_proof_size}
  \end{subfigure}
\hfill
\begin{subfigure}[t]{0.33\textwidth}
    \centering
    \includegraphics[height=4.8cm,width=\textwidth]{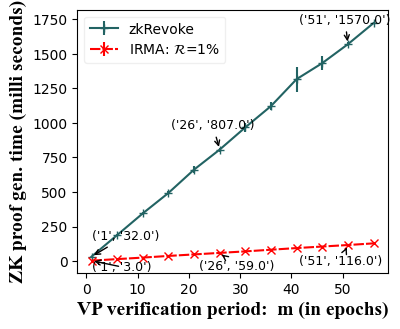}
    \caption{The time to generate proofs for a VP}
    \label{fig:result_zkp_proof_gen_time}
    \end{subfigure}
\hfill
\begin{subfigure}[t]{0.33\textwidth}
    \centering
    \includegraphics[height=4.8cm,width=\textwidth]{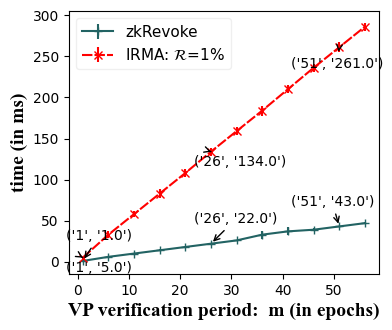}
    \caption{The time to verify proofs for a VP}
    \label{fig:result_zkp_proof_ver_time}
\end{subfigure}

 \caption{The ZK proof size and the time to generate and verify proofs for a single VP having \verificationperiod $m$}
\end{figure*}

\begin{figure*}[t]
     \centering
     \hfill
     \begin{subfigure}[t]{0.4\textwidth}
    \centering
    \includegraphics[height=4.7cm,width=\textwidth]{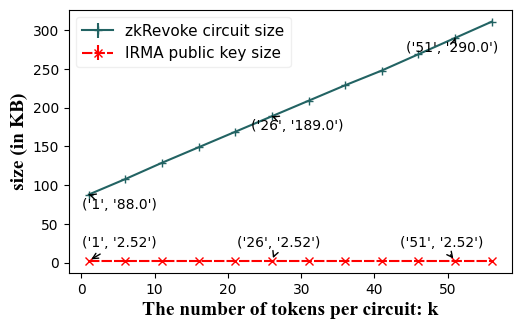}
    \caption{Size: ZK circuits in \name with public keys in IRMA}
    \label{fig:result_circuit_size}
  \end{subfigure}
\hfill
\begin{subfigure}[t]{0.4\textwidth}
    \centering
   \includegraphics[height=4.8cm,width=\textwidth]{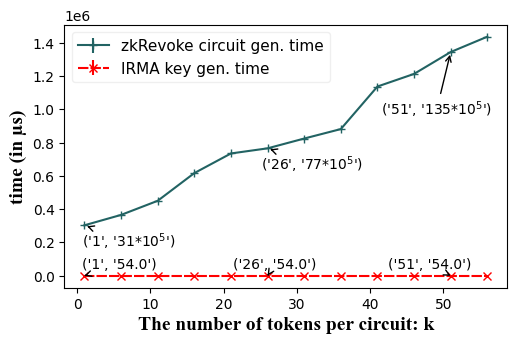}
    \caption{Generation time: ZK circuits in \name vs. public keys in IRMA}
    \label{fig:result_circuit_time}
    \end{subfigure}
    \hfill

 \caption{Comparing the size and generation time of ZK circuits in \name with public keys in IRMA}
\end{figure*}

\subsection{General-purpose ZKPs}
\label{sec:evaluation_zkp}

In our construction, we use general-purpose ZKPs based on the Groth16 ZKP scheme, whereas in IRMA, custom-built ZKPs are used. The objective is to analyze the trade-offs between these approaches and identify the benefits and costs. 

The theoretical analysis of proof size, proof generation time, and proof verification time is discussed, along with the definition of the Groth16 scheme in~\cite{groth2016zkp}. We directly use the scheme as it is, without any modification. Therefore, the analysis also applies to our protocol.
The theoretical analysis of proof size along with the definition of the ZKP scheme used in IRMA is defined in~\cite{baldimtsi2017revocation}. In addition, the proof generation and verification time is also present in~\cite{baldimtsi2017revocation}.

We benchmark the performance of the IRMA and Groth16 schemes here and compare the results. The protocols are benchmarked on the same machine to ensure fair comparison.

 \paragraph*{Metric 6: ZK Proof size.}
The size of a ZK proof in \name is $164$ bytes, whereas the size of a non-revocation proof in \irma is $1831$ bytes, as shown in Figure~\ref{fig:result_zkp_proof_size}.
\name requires $11$x smaller storage compared to \irma.

 \paragraph*{Metric 7: ZK Proof Generation Time.}
Figure~\ref{fig:result_zkp_proof_gen_time} illustrates the differences in the time required to compute proofs in \irma and \name. 
The results show that the proof generation time in IRMA is $10$x faster than the proof generation time in \name.

 \paragraph*{Metric 8: ZK Proof Verification Time.}
Figure~\ref{fig:result_zkp_proof_ver_time} illustrates the differences in the time required to verify all proofs for \m epochs in \irma and \name. 
The results show that the proof verification time in \name is $6$x faster than the proof generation time in \irma.

\paragraph*{Metric 9: ZK circuit.}
To generate and verify proofs, ZK circuits are one of the parameters in \name, whereas public keys are sufficient in \irma. 
In \irma, the ZKP scheme is designed as a digital signature scheme and enables holders and verifiers to use public keys to generate and verify proofs. 
The size of the circuit is impacted only by the extension of proving multiple tokens using a single ZK proof. 
Figure~\ref{fig:result_circuit_size} and Figure~\ref{fig:result_circuit_time} compare the size of the circuits and time to generate them during \setup in \name with key generation time and key size in IRMA. 
The results show that IRMA outperforms \name since public keys are shorter in size and take a significantly smaller amount of time to generate compared to ZK circuits.

\paragraph*{Summary.}
 \name provides shorter proofs ($11\times$ smaller) and shorter verification times ($6\times$ faster) compared to IRMA, reducing bandwidth requirements for holders and computation requirements for verifiers.
However, IRMA provides shorter proof generation times ($10\times$ faster) effecting holders' computation.

Another relevant difference is that circuits generated in \name have a significant size ($\geq 88kB$).
In IRMA, public keys ($2.5kB$) serve a similar role as circuits in \name. 
In \name, verifiers need to retrieve circuits from registry, and holders receive them from issuers and store them. 
Even though the circuit size is larger than public keys used in IRMA, the circuit size is still practical for many use cases since it is in the order of few $KB$.

\section{Conclusions}
\label{sec:conclusion}

In this work, we have proposed \name, a revocation technique for Verifiable Credentials that prevents verifiers from monitoring the revocation status indefinitely. 
\name follows a blacklist approach utilizing time-bound \tokens and Zero-Knowledge Proofs (ZKPs). An issuer computes \tokens corresponding to revoked \VCs after each epoch and inserts those \tokens in a list. 
A holder generates (a) \tokens corresponding to current and future \epochs and (b) corresponding ZK proofs. Afterwards, the holder shares those proofs with a verifier
to prove the correctness of the \tokens. 

\name solves the problem of \textit{untraceability} since verifiers can check the revocation status of each \token only during the corresponding \epoch.
In addition, \name satisfies our auxiliary goals: (a) each holder configures how long a verifier  can verify revocation status corresponding to the holder's VP, (b) a holder shares a ZK proof only once rather than periodically,
and (c) an issuer is not able to learn anything about the verification of VPs.

\bibliographystyle{unsrt}
\bibliography{reference}

\appendix

\section{Untraceability: Security Game}
\label{sec: untraceability_security_game}
The game ($\mathbf{Game}^{untrace}_\mathcal{A}$) is played between a challenger and an adversary $\mathcal{A} = (\mathcal{A}_1, \mathcal{A}_2)$  and consists of four phases: (a) Setup Phase, (b) Query Phase, (c) Challenge Phase, and (d) Guess Phase.
The phases are defined below:

\subsubsection*{Setup Phase}
\begin{enumerate}[noitemsep]
    \item Challenger runs $\mathsf{\Setup}$\,($1^\lambda$) $\rightarrow$ (\sk, \pk, \crs, \tsInit, \durationvar, \RevList, \GS )
    \item Challenger keeps a set $Q_h$ of honest holder's VCs
    \item Challenger initializes query sets $Q_c \leftarrow \phi, Q_r \leftarrow \phi,$ and $Q_p \leftarrow \phi$
    \item Challenger provides (\pk, \crs, \RevList) and the following oracles to Adversary $\mathcal{A}_1$:
    a) \oracle{GenVC} $\rightarrow$ \vc; b) \oracle{Revoke}; $\rightarrow$ \RevListNew or $\bot$; c) \oracle{Refresh} $\rightarrow$ \GS; d) \oracle{GenVP} $\rightarrow$ \vp
\end{enumerate}

\begin{framed}
\noindent\textbf{Oracle \oracle{GenVC}}
\begin{itemize}[noitemsep, topsep=0pt]
    \item \claims $\leftarrow \{0,1\}$
    \item \vc $\leftarrow \mathsf{Issuance}$(\sk, \claims)
    \item $Q_c \leftarrow Q_c \cup \{$vc$\}$
    \item \textbf{return} \vc
\end{itemize}
\end{framed}

\begin{framed}
\noindent\textbf{Oracle \oracle{Revoke}}(\vc, \RevList)
\begin{itemize}[noitemsep, topsep=0pt]
    \item \textbf{if } \vc $\in Q_c$ \textbf{ then}
    \begin{itemize}[noitemsep, topsep=0pt]
        \item \RevListNew $\leftarrow \mathsf{Revocation}$(\vc, \RevList)
        \item $Q_r \leftarrow Q_r \cup \{$vc$\}$
        \item \textbf{return} \RevListNew
    \end{itemize}
    \item \textbf{else return} $\bot$
\end{itemize}
\end{framed}

\begin{framed}
\noindent\textbf{Oracle \oracle{Refresh}}(\ecounter, \RevListNew)
\begin{itemize}[noitemsep, topsep=0pt]
    \item \GS $\leftarrow \mathsf{Refresh}(\mathsf{e}$, \RevListNew)
    \item \textbf{return} \GS
\end{itemize}
\end{framed}

\begin{framed}
	\noindent\textbf{Oracle \oracle{GenVP}}(\pk, \crs, \ecounter, \m, \challenge)
	\begin{itemize}[noitemsep, topsep=0pt]
		\item Fetch a valid \vc, s.t., \vc $\in$  $Q_h \cup Q_c$
		\item vp$_0$ $\leftarrow \mathsf{\Presentation}$\,({\pk, \crs,  \ecounter,  \vc, \m, \challenge})
		\item $Q_p \leftarrow Q_p \cup \{$vp$\}$
		\item \textbf{return} \vp
	\end{itemize}
\end{framed}

\subsubsection*{Query Phase}
Adversary $\mathcal{A}_1$ adaptively queries the oracles \oracle{GenVC},  \oracle{Revoke}, \oracle{Refresh}, and \oracle{GenVP} polynomial many times.

\subsubsection*{Challenge Phase}
\begin{enumerate}[noitemsep]
    \item Adversary $\mathcal{A}_1$ outputs ({\challenge}$_0$, {\challenge}$_1$, $state_{\mathcal{A}_1}$)
    \item Challenger selects two VCs vc$_0$, vc$_1$ where vc$_0$, vc$_1$ $\in Q_h$;  vc$_0$, vc$_1$ $\notin Q_c \cup Q_r$, and  vc$_0$ $\neq$ vc$_1$, and computes:
        \begin{itemize}[noitemsep]
     		\item Verification periods {\m}$_0$, {\m}$_1$
     		\item  The current \epoch value:  \ecounter $\leftarrow$  (\emph{time}()$\ -\ $\tsInit)$/$\durationvar.
            \item vp$_0$ $\leftarrow \mathsf{\Presentation}$\,(\pk, \crs, \ecounter, vc$_0$, {\m}$_0$, {\challenge}$_0$)
            \item vp$_1$ $\leftarrow \mathsf{\Presentation}$\,(\pk, \crs, \ecounter, vc$_1$, {\m}$_1$, {\challenge}$_1$)
        \end{itemize}
     \item After verification period $\max$({\m}$_0$, {\m}$_1$) expires, Challenger samples $b \in \{0,1\}$
     \item Challenger computes the current \epoch value: \ecounter $\leftarrow$  (\emph{time}()$\ -\ $\tsInit)$/$\durationvar and runs the following:
     \begin{itemize}[noitemsep]
         \item  \RevListNew $\leftarrow \mathsf{\Revocation}$\,(vc$_b$, \RevList)
         \item  \GSNew $\leftarrow \mathsf{\Refresh}$\,(\ecounter, \RevListNew) 
     \end{itemize}
     \item Challenger gives (vp$_0$, vp$_1$, \pk, \crs, \GSNew, $state_{\mathcal{A}_1}$) to adversary $\mathcal{A}_2$ 
\end{enumerate}

\subsubsection*{Guess Phase}
\begin{enumerate}[noitemsep]
    \item Adversary $\mathcal{A}_2$ outputs guess $b' \in \{0,1\}$
    \item Game returns 1 if $b = b'$, otherwise returns 0
\end{enumerate}

We consider a hybrid adversary model in which the adversary can act both as a credential holder and as a verifier during the query phase. As a holder, the adversary can request VCs using the \oracle{GenVC} oracle, which returns the full credential, including metadata and signature. As a verifier, it can interact with the \oracle{GenVP} oracle to obtain VPs from honest holders based on arbitrary queries. However, in the challenge phase, the VPs are derived from credentials never revealed to the adversary, reflecting a setting in which honest holders generate proofs using unseen VCs. This models a malicious verifier who has significant observational knowledge, but cannot correlate challenge VPs with previously seen credentials.

\section{Security Analysis}
\label{sec:security}

We prove that \name satisfies {completeness, soundness and untraceability properties}.

\subsection{Completeness}

Our construction satisfies completeness if, for a \VP valid in epoch \ecounter generated from a non-revoked \VC that is not expired in epoch \ecounter, the verification of \VP in epoch \ecounter always succeeds.

\begin{theorem}
If the ZKP system is Complete and the hash function is collision-resistant, then $\Pi$ = (\Setup, \Issuance, \Revocation, \Refresh, \Presentation, \Verification) satisfies the Completeness property. 
\end{theorem}

\begin{proof}[Proof Sketch]
	We prove completeness of \name as follows:
	Let $\mathsf{vp}$ be a valid \VP generated using the \textit{Presentation} procedure in epoch $\mathsf{e'}$ from a non-revoked \vc that is not expired in epoch \ecounter such that $\mathsf{e'} \leq \mathsf{e} \leq (\mathsf{e'}+\mathsf{m})$. The verification of \vp in epoch \ecounter consists of two main components: 
	
\paragraph*{ZK Proof Verification.} For \m tokens, the holder generates \m ZK proofs $\pi_\mathsf{zkp}$. $\Pr$[All ZK Proofs Succeed] = $(1 - {\epsilon}_1)^{\mathsf{m}}$, where ${\epsilon}_1 =  \mathsf{negl}(\lambda)$ due to Groth16 ZKP completeness.
\paragraph*{Revocation Status Verification.} For any valid \epoch \ecounter in the verification period \m, the \textit{VerifyRevocationStatus} algorithm checks if \tokenvar $\in$ \GSNew. Due to the collision resistance of $\mathsf{H}$, for \epoch \ecounter and unique seed values, $\mathsf{H} $({\seed}$_1$, \ecounter) $\neq$ $\mathsf{H} $({\seed}$_2$, \ecounter), meaning it is not possible for any revoked \vc to result in the same \token value. Therefore, for epoch \ecounter: \tokenvar = $\mathsf{H}$(\vc.\seed,  \ecounter) $\notin$ \GSNew: $\Pr$[\tokenvar $\notin$ \GSNew \!\!] $= 1 - {\epsilon}_2$, where ${\epsilon}_2 =  \mathsf{negl}(\lambda)$ due to the collision-resistance property of hash function. Hence, for all \m epochs: $\Pr$[All revocation checks pass] $= (1 - {\epsilon}_2)^{\mathsf{m}}$

Hence, combined probability of verification of \vp :\\
  $\Pr[\mathsf{\Verification (pk, crs, e, blacklist'_{iss} , m, \mathrm{vp}, challenge)} = 1]$ = 
$\Pr$[All ZK Proofs Succeed] $\cdot$ $\Pr$[All revocation checks pass]
	\begin{align*}
		\Pr[\mathsf{Verification}(\mathrm{\vp}, \cdots) = 1] 
		&= (1 - \epsilon_1)^{\mathsf{m}} (1 - \epsilon_2)^{\mathsf{m}} \\
		&\geq 1 - \mathsf{negl}(\lambda)
	\end{align*}

\end{proof}

\subsection{Soundness}
\label{sec:security_soundness}
 Our scheme satisfies soundness if for a \VP which is either invalid or created using a revoked or expired \VC in \epoch \ecounter, the verification of \VP in  \epoch \ecounter results in failure.

\begin{theorem}
If the ZKP system is sound and Zero-Knowledge, the digital signature scheme is Unforgeable, and the hash function is pre-image resistant, then $\Pi$ = (\Setup, \Issuance, \Revocation, \Refresh, \Presentation, \Verification) satisfies the Soundness property.  
\end{theorem}

\begin{proof}[Proof Sketch]
	We prove soundness of \name as follows: 

	Regarding the invalid VPs, a VP can be formed in many ways. Nevertheless, if the VP does not include correct information, if the ZKP system is sound and the digital signature scheme is unforgeable, then the verification of the ZK proof included in the VP would result in failure. 
	
	Regarding a \VP generated from revoked or expired \VC using the presentation procedure, the \textit{VerifyRevocationStatus} algorithm deterministically captures the revocation status of \tokens corresponding to the \VP in epoch \ecounter and checks for expiry.

	We justify the soundness property of $\Pi$ by analyzing two attack scenarios. 
    Let $\mathcal{A}$ be a malicious holder of a VC \vc, and let \vp be a VP generated from \vc by presentation. The goal for $\mathcal{A}$ is to create a VP $\mathsf{vp}'$ that differs \vp by one of the following:
	(a) $\mathsf{vp}'$ includes a different \tokenvar than \vp, avoiding revocation checks, (b) $\mathsf{vp}'$ includes different claims than \vp, or (c) $\mathsf{vp}'$ includes a different expiration period than \vp, avoiding expiration checks.

    \paragraph*{Scenario 1: Generating a proof.} The ZK circuit verifies the generation of \tokenvar from \seed and \ecounter.
	$\mathcal{A}$ may attempt to generate $\mathsf{vp}'$ by selecting a different \seed, changing the claims or the expiration period. However, the ZK circuit verifies that these values are signed by the issuer. The adversary $\mathcal{A}$ can try to forge the signature and proof. However,
    \begin{asparadesc}
    	\item[Signature Forgery: ] $\Pr$[$\mathcal{A}$ forges valid signature on $s'$] $\leq \mathsf{negl}(\lambda)$ due to signature unforgeability.
    	\item[ZK Proof Forgery: ] $\Pr$[$\mathcal{A}$ generates valid $\pi_\mathsf{zkp}$ without proper signature] $\leq \mathsf{negl}(\lambda)$ due to ZKP Soundness.
    	\item[Combined: ] $\Pr$[Scenario 1 succeeds] $\leq \mathsf{negl}(\lambda)$
    \end{asparadesc}

	\paragraph*{Scenario 2: Reusing a proof.} Suppose $\mathcal{A}$ obtains a valid VP $\mathsf{vp}'$, that includes the \tokenvar, and the ZK proof $\pi_{\mathsf{zkp}}$. 
	$\mathcal{A}$ may try to either reuse the proofs instead of generating new ones or recover private inputs used in the proofs such as digital signatures and seed values. However, the reused proof will not verify with different claims, expiration period, or most importantly, a different challenge. In addition, recovery of private inputs is not computationally feasible.
	\begin{asparadesc}
    	\item[ZK Proof Soundness: ] $\Pr$[$\pi_\mathsf{zkp}$ verifies with different public values] $\leq \mathsf{negl}(\lambda)$ due to ZKP Soundness.
    	\item[Recovery from ZK Proofs: ] $\Pr$[$\mathcal{A}$  learns signature and other private inputs from $\pi_{\mathsf{zkp}}$  $\leq \mathsf{negl}(\lambda)$ due to ZKP zero-knowledge.
    \end{asparadesc}

	Therefore, $\mathcal{A}$'s best strategy has success probability:  $\Pr$[$\mathcal{A}$ constructs passing \vp\!\!\!]
  	  $\leq \max(\mathsf{negl}(\lambda), \mathsf{negl}(\lambda)) =  \mathsf{negl}(\lambda)$.
  	  Hence;
  	\begin{align*}
		& \Pr[\mathsf{\Verification (pk, crs, e, blacklist'_{iss} , m, \mathrm{vp}, challenge)} = 0] \\
  		&= 1 - \Pr[\mathsf{Verification}(\mathrm{\vp}, \cdots) = 1] \\
  		&\geq 1 - \Pr[\mathcal{A} \text{ constructs passing } \mathrm{\vp}] \\
  		&\geq 1 - \text{negl}(\lambda)
  	\end{align*}

\end{proof}

\subsection{Untraceability}

\begin{theorem}
If the ZKP system is Zero-Knowledge and the hash function is pre-image resistant, then $\Pi$ = (\Setup, \Issuance, \Revocation, \Refresh, \Presentation, \Verification) satisfies the Untraceability property. 
\end{theorem}
\begin{proof}[Proof Sketch]
We prove untraceability by showing that an adversary $\mathcal{A} = (\mathcal{A}_1, \mathcal{A}_2)$ has at most negligible advantage in the untraceability game defined in Definition~\ref{def:untraceability}. 
In the challenge phase of the game, the adversary $\mathcal{A}_2$ receives (vp$_0$, vp$_1$, \pk, \crs, \GSNew, $state_{\mathcal{A}_1}$). Tokens for revoked vc$_b$ appear in \GSNew.  The goal of $\mathcal{A}_2$ is to distinguish whether vc$_0$ or vc$_1$ was revoked. In other words, $\mathcal{A}_2$ needs to track the revocation status of the vc$_{b}$ corresponding to vp$_{b}$ after its \verificationperiod is over. 
Below we show that this is not feasible, since $\mathcal{A}_2$ cannot extract the \seed from the tokens or ZK proofs in the VPs, and therefore cannot link \tokens from different \epochs.

\paragraph*{Scenario 1 : Recovering the seed from tokens.}  
$\mathcal{A}_2$ might attempt to recover the seed from \tokens in the VPs, then compute the current \epoch token to check against \GSNew. 
For any \tokenvari{i} $\in$ vp$_b$.\tokenvars, we have \tokenvari{i} = $\mathsf{H}$(vc$_b$.\seed, \ecounteri{i}) for \epoch \ecounteri{i}. 
By the pre-image resistance of $\mathsf{H}$:
$\Pr$[$\mathcal{A}_2$ recovers vc$_b$.\seed from \tokenvari{i}] $\leq \mathsf{negl}(\lambda)$

\paragraph*{Scenario 2 : Extracting the seed from ZK proofs.}
 $\mathcal{A}_2$ might attempt to extract the \seed from the ZK proofs $\pi_{zkp}$ in the VPs. 
 Each $\pi  \in \pi_{zkp}$  proves knowledge of  vc$_b$.\seed  without revealing it.
By the zero-knowledge property of the ZKP system:
$\Pr$[$\mathcal{A}_2$ learns vc$_b$.\seed from $\pi_\mathsf{zkp}$] $\leq \mathsf{negl}(\lambda)$

\paragraph*{Scenario 3 : Linking tokens across epochs.} $\mathcal{A}_2$ might try to link \tokens from the VPs to\ tokens in \GSNew through correlation attacks. However, \tokens for different epochs are computationally independent:
 (a) \tokenvars in VPs: $\mathsf{H}$(vc$_b$.\seed, \ecounteri{i}) for epochs  \ecounteri{i}  $\in \{$\ecounter, \ecounter +1, \ldots, \ecounter+ {\m}$_b$ $- 1$$\}$, and (b) \tokenvars in \GSNew: $\mathsf{H}$(vc$_b$.\seed, \ecounter') for current epoch \ecounter'  > \ecounter + $\max$({\m}$_0$, {\m}$_1$) $- 1$. 

Since the verification periods have expired ( \ecounter'  > \ecounter + $\max$({\m}$_0$, {\m}$_1$) $- 1$ ), there is a ``gap'' between the epochs covered by VP \tokenvars and the current epoch.
By pre-image resistance and the pseudorandom properties of H, these  \tokenvars appear independent. Therefore, 
$\Pr$[$\mathcal{A}_2$ links \tokenvars across epochs]  $\leq \mathsf{negl}(\lambda)$.

\noindent Combining all attack vectors, 
$\Pr$[$\mathcal{A}_2$ succeeds] $\leq$ $\Pr$[$\mathcal{A}_2$ recovers vc$_b$.\seed from \tokenvari{i}]  + $\Pr$[$\mathcal{A}_2$ learns vc$_b$.\seed from $\pi_\mathsf{zkp}$] + $\Pr$[$\mathcal{A}_2$ links \tokenvars across epochs]
$\leq \mathsf{negl}(\lambda) + \mathsf{negl}(\lambda) + \mathsf{negl}(\lambda) \leq \mathsf{negl}(\lambda)$.
Therefore;
\begin{displaymath}
	| \Pr[\mathbf{Game}^{untrace}_\mathcal{A}\, (1^\lambda) = 1 ] - 1/2 | \leq \mathsf{negl}(\lambda)
\end{displaymath}
	
\end{proof}

\section{\name: Microbenchmarks}

\begin{table}[th]
\caption{The Cost of Storing Tokens in Smart Contract}
\label{tab:token_cost}
\begin{center}

\begin{tabular}{m{3cm}m{1.5cm}m{1.5cm}}
\toprule
\textbf{Number of Tokens} &  \textbf{Size} & \textbf{Gas} \\
\midrule
$1$ & $32$ B & $21993$ \\
\midrule
$10$ & $320$ B & $26591$ \\
\midrule
$100$ & $3200$ B & $72443$ \\
\midrule
$1000$ & $32$ KB & $553240$ \\
\midrule
$10000$ & $320$ KB & $5339501$ \\
\midrule
$100000$ & $3200$ KB & $53201326$ \\
\bottomrule
\end{tabular}
\end{center}
\end{table}

Table~\ref{tab:token_cost} presents the cost of storing \tokens in the Ethereum Blockchain.
Ethereum blockchain network limits the maximum gas used by a transaction to $30$ million~\cite{eip7825} units. Due to this limitation, it is possible to store only $1200$ \tokens per block in the smart contract since it requires close to $30$ million units of gas, the allowed gas limit per block. Thus, storing more than $1200$ \tokens would require sending multiple transactions, each requiring a new block. Storing a single \token (32 bytes) consumes $21993$ units of gas.

Our ZK circuit consists of $9315$ constraints when $k=1$. The size of our circuit is $89667$ bytes. Similar to the limit on the number of tokens stored, storing the complete ZK circuit would require multiple transactions.

\end{document}